\documentclass[annual]{acmsiggraph}
\pdfoutput=1

\TOGonlineid{0146}
\TOGvolume{0}
\TOGnumber{0}
\TOGarticleDOI{1111111.2222222}
\TOGprojectURL{}
\TOGvideoURL{}
\TOGdataURL{}
\TOGcodeURL{}

\usepackage{color}
\usepackage{soul}
\usepackage{multirow}
\usepackage{rotating}
\definecolor{turquoise}{cmyk}{0.65,0,0.1,0.1}
\definecolor{purple}{rgb}{0.65,0,0.65}
\definecolor{darkgreen}{rgb}{0.0, 0.5, 0.0}
\definecolor{darkred}{rgb}{0.5, 0.0, 0.0}
\definecolor{darkblue}{rgb}{0.0, 0.0, 0.5}
\definecolor{blue}{rgb}{0.0, 0.0, 1.0}

\newcommand{\changed}[1]{{#1}}
\newcommand{\erase}[1]{}
\newcommand{\tog}[1]{{#1}}
\newcommand{\togerase}[1]{}

\newcommand{\hide}[1]{{}}

\newcommand{\eg}{{\textit{e.g., }}}
\newcommand{\ie}{{\textit{i.e., }}}

\usepackage{amsmath}
\usepackage{bm}

\title{Understanding and Exploiting Object Interaction Landscapes}

\author{
S\"oren Pirk$^{1}$\thanks{e-mail:soeren.pirk@gmail.com} \hspace{3mm}
Vojtech Krs$^{2}$\hspace{3mm}
Kaimo Hu$^{2}$ \hspace{3mm}
Suren Deepak Rajasekaran$^{2}$ \hspace{3mm}
Hao Kang$^{2}$ \hspace{3mm}\\
Bedrich Benes$^{2}$ \hspace{3mm}
Yusuke Yoshiyasu$^{3}$ \hspace{3mm}
Leonidas J. Guibas$^{1}$\\[2mm]
$^1$Stanford University, USA\hspace{3mm}
$^2$Purdue University, USA\hspace{3mm}
$^3$CNRS-AIST JRL, Japan
}
\pdfauthor{S\"oren Pirk"}


\begin{document}
\teaser{
 \includegraphics[width=\linewidth]{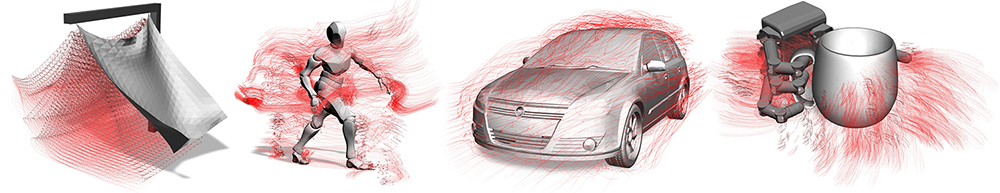}
 \caption{Different interaction landscapes representing the interactions of a motion driver with a static object. We capture the motion trajectories (red) and encode their signatures into a descriptor that can be used for comparing interactions. From left to right: a cloth simulation interacting with a support structure, a human model walking on a floor, a wind simulation interacting with a car, and a robotic hand grasping a cup.}
 \vspace{-2mm}
\label{fig:teaser}
}

\maketitle

\begin{abstract}
Interactions play a key role in understanding objects and scenes, for both virtual and real world agents.\hide{Understanding the function of shapes is an important open problem in computer graphics.} We introduce a new general representation for proximal interactions among physical objects that is agnostic to the type of objects or interaction involved. The representation is based on tracking particles on one of the participating objects and then observing them with sensors appropriately placed in the interaction volume or on the interaction surfaces. We show how to factorize these interaction descriptors and project them into a particular participating object so as to obtain a new functional descriptor for that object, its \emph{interaction landscape}, capturing its observed use in a spatio-temporal framework. Interaction landscapes are independent of the particular interaction and capture subtle dynamic effects in how objects move and behave when in functional use. Our method relates objects based on their function, establishes correspondences between shapes based on functional key points and regions, and retrieves peer and partner objects with respect to an interaction.
\end{abstract}





\section{Introduction}

The goal of this work is to define a general-purpose descriptor for interactions among physical objects (animate or inanimate) and/or elements (air, water, fire), such as a human sitting on a chair, grasping a cup, or playing a musical instrument, as well as  water pouring into a glass, air flowing past a driving car, etc. Such interactions usually involve pairs or multiple objects in relative proximity or contact. Current representations of interactions are polarized at two extremes. On the one hand, there are very high level interaction descriptions like those in natural language ("John grasped his mug and poured coffee into it"); computer vision has expended a lot of effort on mapping images and videos to such descriptions (action recognition). On the other hand, there are the detailed simulations performed in computer graphics requiring full representation of the geometry of the environment, \eg a CAD model of a mug, finite element models for deformable objects like human fingers grasping the mug, as well as material properties such as viscosity for the fluid (coffee) flowing into the mug. In our work we aim for an {\em intermediate-level representation}, capturing information about the geometry and physics of the interaction than just a natural language description, yet abstracted away from all the details tied to the specific geometry and physics discretizations used in the simulation. Our motivation is to compare and understand the similarity between the actions of a human sitting on a chair or on a bed, while being able to distinguish the actions of a human hand grasping a mug from its handle, as opposed to from its rim (because perhaps there is no handle and the mug is too hot) --- and to do so in a way compatible with multiple virtual realizations of these action scenarios as well as with multiple capture technologies.

We believe that the types of actions that an object participates in are crucial in capturing the semantic identity of the object --- as much as its geometry, appearance, materials, etc., quantities that have been traditionally studied in computer graphics. We want to understand in a more computational form the old notion that form and function are deeply inter-correlated. Towards this end, we study how to project actions and interactions onto each of the participating objects or media, so that an object's interactions become a new signature for the object in a way that captures the essential spatio-temporal aspects or patterns of the interaction {\em from the point of view of the object}. Our ultimate goal is, given an object, to summarize all the interactions it has experienced into one unified mathematical representation, and to be able to use such interaction landscapes to retrieve similar objects, to recover partner objects that fit a particular interaction (given a screw model, find the appropriate screwdriver), or to infer functionality of a new object.


Understanding shape functionality remains one of the key challenges in shape analysis and geometric modeling. Researchers have not only studied the perception of shapes~\cite{FeaturesOfSimilarity.pdf,ashby1992multidimensional}, but have also analyzed their structure, form, and similarity with a wide variety of computational models and representations~\cite{Mitra:2013:SSP:2542266.2542267,CGF:CGF12734}. It has been recognized that shape modeling is a fundamental problem that must include a high-level understanding of the semantic meaning of a shape --- one of the key observations of previous studies is that a shape is not purely described by its form but also by its function~\cite{Hu:2015:ICT:2809654.2766914,Hu:2016:LOF:2897824.2925870}. Embedding a shape in its interaction space enables a more thorough understanding of its functionality, a key requirement in many application areas, \eg shape correspondence and similarity, reconstruction and content creation.

Some of the previous approaches formulate the problem of understanding shapes by looking at human-centric interactions to learn shape affordances \hide{. Fitting a human agent to a shape implicitly reveals information about its affordance}~\cite{Kim:2014:SHS:2601097.2601117}. Moreover, capturing real humans allows us to learn the function of an environment and to infer possible interactions for unobserved scenes~\cite{Savva:2014:SIA:2661229.2661230}. Others analyze shapes by defining functionality models or by using distinct labels for classifying interactions~\cite{Sutton19941743,Laga:2013:GCS:2516971.2516975}. More recently, it has also been recognized that spatial relations can infer a context-aware signature of object-object interactions~\cite{Hu:2015:ICT:2809654.2766914} that even allow the analysis of shapes~\cite{Hu:2016:LOF:2897824.2925870}. The diversity of these contributions illustrates the breadth of the interaction modeling area.

Our quest for a universal interaction descriptor is made challenging by several factors. First, there is a great deal of interaction diversity, as the spatio-temporal motions involved within a particular interaction class (\eg using a mug) can be very diverse, depending on the exact geometry of the objects, their materials (solid vs.~fluid), state (empty vs.~full), etc. At the same time, we require a robust distance metric or a way to establish correspondences between such motions in a way that captures their essential interaction qualities, while not being overly affected by irrelevant details. Our approach towards this goal is to look at the trajectories of motion particles sampled from a partner object, as they move in the vicinity of the object of interest. We call this partner object the \emph{motion driver}, as it generates these particle motions. In our coffee mug example, we look at particles on  the human hand grasping the mug, or particles in the coffee fluid pouring into the mug. To observe the partner motion particle trajectories, we establish a set of \emph{sensors} on or near the object of interest, each of which is tracking particle motions in a region of free space near it. \tog{The sensors then build a descriptor of the trajectories observed in the part of space each is responsible for --- and altogether these local descriptors form our interaction signature. \togerase{, which we call an \emph{interaction landscape}}}


Unlike our emphasis on both the temporal and the spatial aspects of the interaction, most of the previous approaches encode interactions as static relationships and they focus either on specific object categories (\eg man-made shapes) or only explore the interactions of shapes based on specific agents (\eg human-centric). Our interaction landscapes are more general, allowing us to relate objects with their peers (others like them) as well as with their interaction partners. Furthermore, while numerous prior studies have analyzed object surfaces and their properties, our work \erase{for the first time}puts the focus not just on the surfaces of objects but also on the space around the objects where much of the interaction with partners happens --- as well as on the temporal evolution of activity in these regions. \hide{While this paper deals primarily with virtualized interactions, we envision a follow up project, where we capture such descriptors directly, by observing interactions in the real world and tracking points of interest in the environment.}


We perform a number of experiments to demonstrate the relevance of our interaction descriptor for analyzing the functionality of shapes. In particular, we show that it is well-suited for shape correspondence and classification tasks. Additionally, we show that our approach can capture and encode multiple functions of an object, \ie a cup can be grasped at the top for carrying it, while it can also be grasped from the side for drinking. Figure~\ref{fig:teaser} shows examples of four interaction landscapes, each representing the interaction of a motion driver with a static object. We capture the motion trajectories (shown in red) and encode their signatures with a novel interaction encoding method. In summary, we claim the following contributions:
\vspace{3mm}
\begin{itemize}
\item We introduce a novel representation for proximal interactions between physical objects and a flexible framework that allows an exploration of the interaction dynamics driven by input data from various sources, \eg simulations, animations, motion capture, and \changed{RGB-D scans}.
\item We show how to factorize interactions into descriptors for each of the participating objects and how to compare such descriptors.~The resulting interaction landscapes establish correspondences between shapes based on functionally-defined keypoints or regions.
\item{We evaluate our method against state-of-the-art descriptors such as \changed{the Light-field Descriptor (LFD), the Intersection Bisector Surface (IBS) and the Interaction Context (ICON)}, and show the relevance of our descriptor for applications, such as shape retrieval and saliency estimation.}
\end{itemize}

\hide{This in turns allows us to organize and search objects according to the similarities of their interactions so that, given a particular object, we are able to retrieve others that are similar or complementary to it.}
\section{Related Work}
Shape completion and segmentation, similarity assessment of shapes, and structure-aware modeling have been recognized as fundamental problems in a variety of application domains~\cite{Mitra:2013:SSP:2542266.2542267}.
Traditionally, many of the previous methods in these areas rely on geometric descriptors to encode local features of shapes and forms. By employing functions on the geometry, these methods provide more discriminative attributes for relating shapes, often formulated with the goal of being robust or invariant against specific transformations~\cite{CGF:CGF12734}. Functions on shapes have also been the basis for the functional formulation of maps and correspondences between shapes~\cite{fmapssig2012}. More recently, efforts have shifted towards encoding the structural meaning of shape parts and high-level semantic information. \changed{Structure-aware modeling of shapes requires knowledge of symmetries~\cite{journals/cgf/MitraPWC13,Tevs:2014:RSV:2601097.2601220}, similarities of parts~\cite{Sidi:2011:UCS:2070781.2024160}, variability~\cite{CGF:CGF12039} and the object's embedding in its context~\cite{Zhao:2014:ISU:2631978.2574860,Hu:2015:ICT:2809654.2766914}}.

Analyzing the uses and affordances of a model provides a deeper understanding of shapes. While early attempts explicitly model the functionality of shapes in the context of specific categories~\cite{Sutton19941743} or decompose models into sets of volumetric functional shape primitives~\cite{323839}, representing functionality allows the resolution of object-functionality correspondences, even for high variance data sets. Pechuk et al.~\shortcite{Pechuk:2008:LFO:1363359.1363379} derive functionality labels by employing a multi-level hierarchy for the automated classification of parts. Laga et al.~\shortcite{Laga:2013:GCS:2516971.2516975} encode the context of a model as structural relationship. Their approach models shapes as interconnected parts in order to derive pairwise-correspondences of 3D shapes and to learn the functionality of shape components through supervised classification. Zhu et al. \shortcite{Zhu2014} propose an indirect learning approach of poses through training a knowledge base.

Several approaches exist to explore interactions of human agents with shapes or even entire scenes. Bar-Aviv et al.~\shortcite{Bar-AvivR06} and Liu et al.~\shortcite{Liu2015110} propose agent-centric systems that analyze the performance of human avatars for classification and interaction tasks. While Grabner et al.~\shortcite{5995327} only concentrate on analyzing the functionality of chairs, Kim et al.~\shortcite{Kim:2014:SHS:2601097.2601117} detect functional similarities by predicting human poses based on a trained affordance model. The human agent can also be observed from images~\cite{Chao2015} and videos~\cite{Gupta:2009:OHI:1608576.1608766}. \changed{Li et al.~\shortcite{4293017} and Song et al.~\shortcite{6130360} explicitly limit themselves to investigating visual grasp affordances from 2D measurements.}

In SceneGrok, Savva et al.~\shortcite{Savva:2014:SIA:2661229.2661230} observe the behavior of real humans interacting with an environment. They capture RGB-D data to infer action maps and train a classifier that allows the transport of the interaction knowledge to other, unseen environments. \changed{In a more recent work Savva et al.~\shortcite{Savva:2016} show that it is possible to infer higher level semantic meaning from observed human-centric interactions}. \erase{More recently} Knowledge about interactions can \erase{even}be employed as a means for reconstructing objects and scenes. Fisher et al.~\shortcite{Fisher:2015:ASS:2816795.2818057} used virtual agents to associate object arrangements with typical activities to derive more plausible scene setups. Tzionas et al.~\shortcite{Tzionas2015} introduced a 3D reconstruction pipeline based on how hands interact with a target object, which even works for featureless or symmetric shapes. \changed{Instead of focusing on the interaction of real humans with their environment, other works concentrate on encoding dynamic motions of virtual humanoid agents for retargeting and motion adaptation purposes~\cite{Ho:2010:SRP:1778765.1778770,Al-Asqhar:2013:RDI:2485895.2485905}. }

In contrast to the previous methods, we are interested in exploring interaction partners --- complementary shapes that accomplish a function jointly with the given shape.\hide{For example, a screwdriver interacting with a screw, a spoon stirring a fluid, or a human sitting down on a chair.} This notion is similar to what Cain~\shortcite{351268} proposed as \textit{Motion Constraints}; however, our method focuses on capturing the dynamic nature of interactions of two or more shapes and not just static relationships. Moreover, we show that interactions can be factorized by analyzing their motion patterns in an appropriate vector space~\cite{Liefei2011}. Related mathematical ideas based on ``currents'' for describing surfaces can also be found in Durlemann~\shortcite{Durleman2010}.

\erase{Finally, o} \changed{O}ur approach is similar to the \textit{Shape Flow} method proposed by Jiang and Martin \shortcite{jiang2008}, who detect and encode flow lines in videos and the \textit{HON4D} descriptor introduced by Oreifej and Liu~\shortcite{6618942} that encodes interactions captured in RGB-D data. Unlike the previous specialized approaches, we propose a \tog{more} general framework that allows us to encode a wide variety of different interactions in 3D setups. Additionally, we show that our signatures show similar characteristics to the recently proposed ICON descriptor \cite{Hu:2015:ICT:2809654.2766914} for static scenes, which makes them well suited for tasks such as estimating shape correspondences and assessing shape similarity.

We finally note that our novel particle-based representation has multiple advantages over prior interaction encodings. It is independent of the specific geometric and physical model used to obtain the interaction simulation. Moreover, trajectory collections are nice mathematical objects on which a variety of analysis tools, such as vector fields, flow analysis, etc., can be applied.

\section{Overview}
The objective of our work is to encode and abstract the dynamic characteristics of interactions. More specifically, we focus on interactions of one object in motion, which we refer to as \textit{motion driver}, with another static object (\eg a chair or a cup) that is acting as the action receiver. The motion driver can be any source of motion; to explore the versatility of our approach, we have employed motions from a variety of sources, such as simulators specialized for grasping or fluids, character animations, physics-based motions, and motion captured data.

\begin{figure}[t]
\includegraphics[width=\linewidth]{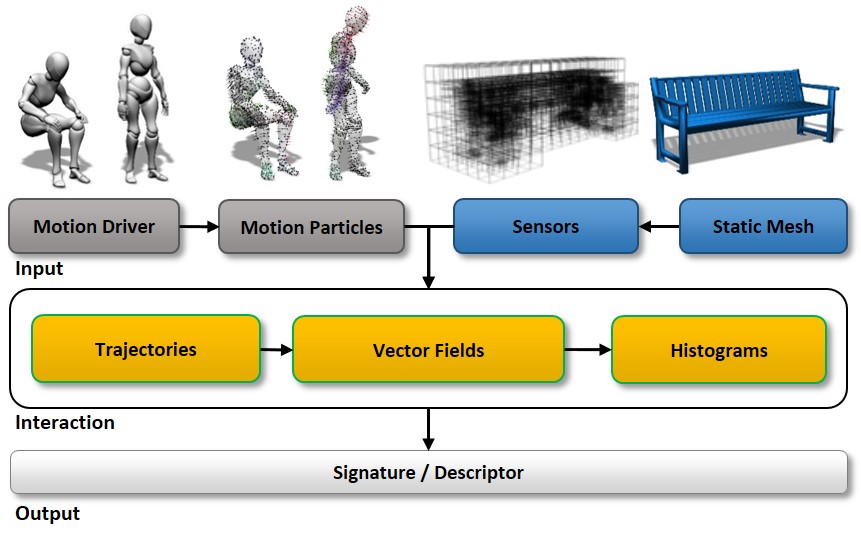}\hfill
\vspace{-2mm}
\caption{Overview of the pipeline for computing interaction landscapes. Given an input mesh and a motion driver, we track the movement of motion particles attached to the moving object in local sensor regions. In each region, the sampled trajectories of the motion particles are converted into vector fields that are analyzed for unique patterns in the motion flow. By quantizing the vector space we derive a global signature that allows us to embed objects in their interaction space and to relate them based on their functionality.}
\label{fig:overview}
\end{figure}

The surface of the motion driver (see Figure~\ref{fig:overview}) is covered with \textit{motion particles} that we use to track the motion. For particle-based simulations, such as Lagrangian fluids, we randomly select a subset of the actual simulation particles to track. Depending on the type of motion that causes the interaction, we provide the system with a number of parameters, such as the location and orientation of an emitter, the number of particles for a fluid flow, or the number and properties of transformations for animations and motion-capture data sets.

\begin{figure*}[t]
\includegraphics[width=\linewidth]{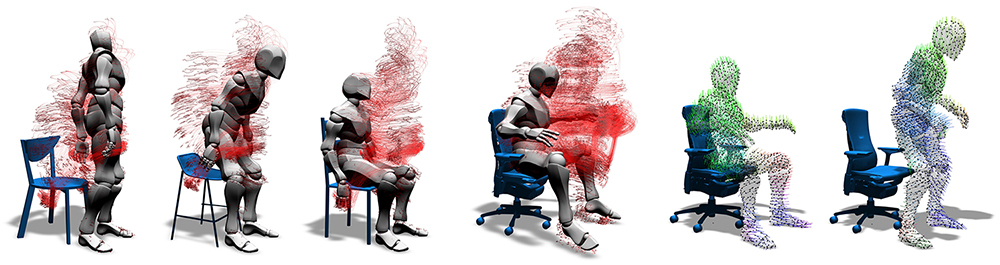}\hfill
\vspace{-2mm}
\hspace*{11mm}(a)\hspace*{22mm}(b)\hspace*{24mm}(c)\hspace*{27mm}(d)\hspace*{30mm}(e)\hspace*{28mm}(f)\hfill~
\caption{A human model sitting down on various chairs. Depending on the performed action, the motion trajectories, shown in red, can vary: compare (a) and (d). We capture the trajectories by uniformly distributing motion particles on the surface of the motion driver (human model). We record the direction and speed of the movement of each motion particle: (d), (e), and aggregate their trajectories into signatures.}
\label{fig:agent_placement}
\vspace{-3mm}
\end{figure*}

Our approach is based on the observation that the interaction of two shapes defines a unique pattern that characterizes their form and function. To encode the dynamics of an interaction we examine how a motion particle approaches an object. We sample the surface of the input shape to distribute sensors that track motion particles passing by, with each sensor measuring a region on or near the surface. These regions span an interaction space that we refer to as the \textit{interaction landscape}. As it is not known where the interactions are likely to happen beforehand, we spatially subdivide the potential interaction space based on the input geometry. Every motion particle that enters one of the sensor regions creates a unique motion trajectory that characterizes the occurring interaction.

We transform the captured trajectories into local vector fields for each of the sensors. Operating on vector fields allows us to analyze the properties of the observed interaction using powerful mathematical tools. More specifically, we detect a number of attributes of the vector fields, such as the vector and tensor magnitude, dilation and vorticity, as well as particle velocity and orientation. We build histograms over these attributes to factorize the interaction and capture the response of different shapes to the same motion driver by computing their relative distance over a set of histograms.

\section{Interaction Landscapes}\label{sec:landscape}
An interaction landscape is a 4-dimensional (three spatial and one temporal) interaction space of two shapes participating in an interaction. The input to our algorithm is a \changed{2-dimensional} surface mesh~$M$ of the observed static shape\changed{ that can even be a triangle soup} and a time variant motion driver~$D$ (the partner moving object or medium)\changed{, both embedded in ${\rm I\!R}^3$}. The motion driver encodes the source of movement that initiates the interaction; for our tests we used moving surfaces and simulations. \changed{We define an interaction landscape by specifying a set of sensors $S=\{s_1,...,s_n\}$ distributed in the \erase{outer domain} interaction space} of~$M$ and motion particles~$A=\{a_1,...,a_m\}$ originating from~$D$. \changed{The interaction space~$U$ is a fixed size bounding box with \tog{manually added extra space around the object}, large enough to encompass all objects of a particular class.} Depending on the type of motion, motion particles are either stochastically sampled on the surface (\eg for animations), or directly advected by the simulation (\eg for fluids). \changed{Moreover, we expect shapes of the same class to share a common orientation and require a canonical arrangement of the interaction landscape with respect to the observed shape.}

\subsection{Motion Drivers}
\label{sec:motion_driver}
Our method provides a \tog{more} general means for exploring different types of interactions represented by the motions of a motion driver. To capture the variance of motions of shape parts relative to each other and against the observed shape, we compute a distribution of motion particles on the surface of the motion driver \changed{that covers the entire surface but with an emphasis on salient regions.}

\erase{For moving surfaces} \changed{We ensure this distribution of samples on the surface of~$D$ by employing a bilateral farthest point sampling strategy.} Given a motion driver~$D$, we first initialize the particle set with a single position entry~$P=\{p_1\}$, with~$p_1$ being stochastically sampled on~$D$. \tog{We iteratively add the farthest point $s$ on $D$ to $P$ until a specified number of particles is distributed.} To capture salient features of $D$, we use a bilateral distance measure, that enlarges distances in\erase{dense} \changed{regions of the surface with finer details} and thereby emphasizes the sampling in these regions~\cite{Chen:2013:BBN:2508363.2508375}. \tog{\changed{We add new samples to $P$ by computing}\erase{We compute} a density function~$\rho$ over all mesh vertices based on their \textit{local feature sizes (lfs)} and define the distance measure $\bar{d}(s,p)$ as
\begin{equation}
\label{equ:16}
\bar{d}(s,p) = d(s,p) \cdot \rho,~\rho(p) = 1/\mathrm{lfs}(p)^2
\end{equation}
\changed{for each existing sample $p \in P$,} where $s$ is a particle sample candidate, $d(s,p)$ the Euclidean distance between $s$ and $p$ in \changed{${\rm I\!R}^3$}, and $\mathrm{lfs}(p)$ the inverse density function that measures the saliency of $p$ \changed{based on its distance to the medial axis of the mesh}~\cite{Amenta:1998:NVS:280814.280947}. Vertices of the mesh are denoted as $s$ and $p$, and the sample candidates $s$ are selected from the remaining set of vertices of the mesh. For each $p_i$ in $P$ we compute the distance according to~Eq.~\ref{equ:16}. This definition is not symmetric as we only consider the saliency of the points that already exist in $P$. The $\mathrm{lfs}$ is the distance to the medial axis of the mesh~\cite{Li:2015:QCM:2870647.2753755}, which is insensitive to its resolution. Edges smaller than a certain threshold are not considered. Figure~\ref{fig:agent_placement}~(e) and (f) illustrate the resulting farthest point sampling for an animated surface mesh.
}

For fluid flows and other Lagrangian simulations, positions for motion particles are directly derived from the simulation. In our implementation, fluid particles are randomly initiated. Hence, we can stochastically sample a subset of motion particles from the set of fluid particles. We track the selected motion particles until the life time of the corresponding fluid particle has ended and initiate a new particle at the fluid emitter. Other implementations of fluid simulations might require more refined sampling strategies to assure complete coverage of particle positions across the motion driver. For most of our tests we used a few hundred to a couple of thousand particles, for both moving surfaces and fluid flows.

\begin{figure}[t]
\vspace{0.2cm}
\includegraphics[width=\columnwidth]{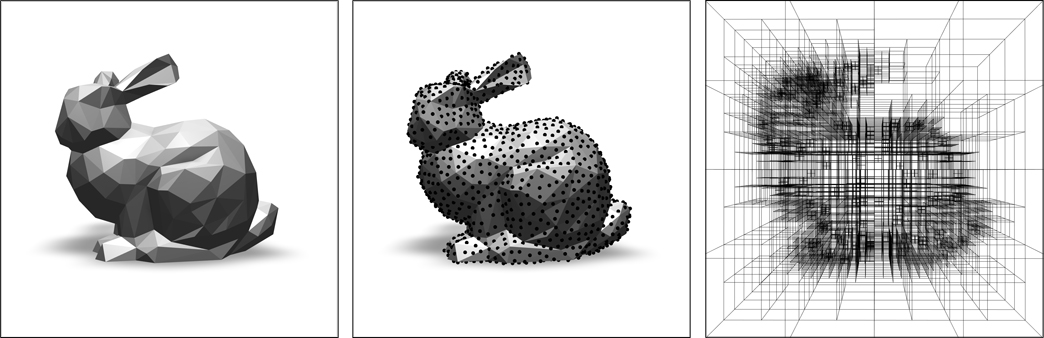}\hfill
\caption{\changed{To capture interactions we define sensor regions in the domain around the observed shape (left). We compute samples on its surface (middle) and use them to produce a spatial subdivision of a graded set of cubes that get finer close to its surface. The location and size of each sensor corresponds to the leaf nodes of an Octree that generates the sensor regions~(right).}}
\label{fig:sensor_placement}
\vspace{-2mm}
\end{figure}

\begin{figure*}[t]
\includegraphics[width=\linewidth]{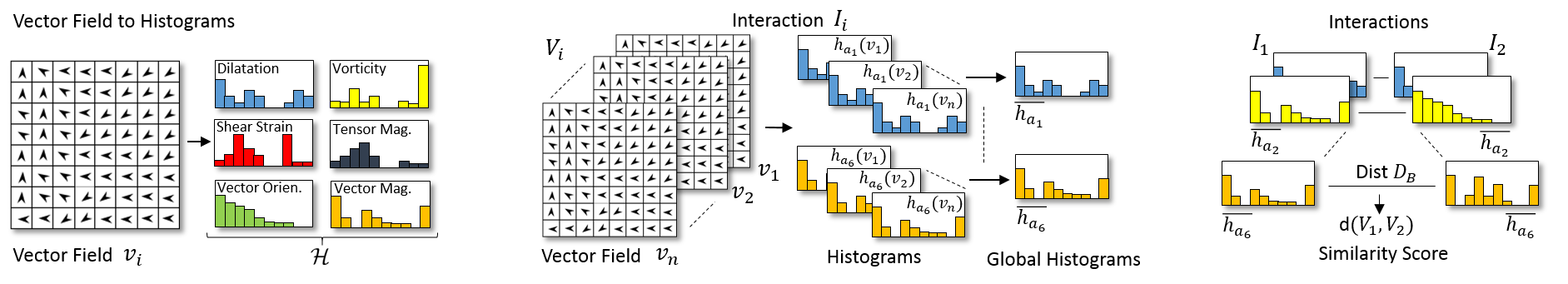}\hfill
\vspace{-2mm}
\hspace*{26mm}(a)\hspace*{67mm}(b)\hspace*{55mm}(c)\hfill~
\vspace{-2mm}
\caption{A 2D illustration of our descriptor: each sensor region is converted into a vector field that allows us to derive a set of six attributes describing patterns in the captured motion flows,  which we quantize by computing histograms for each attribute~(a). We average the local attribute histograms over all vector fields to produce a global histogram for each attribute~(b). Finally, we define a similarity score between two interactions by computing the sum of relative distances between two corresponding global histograms~(c).}
\vspace{-4mm}
\label{fig:flow_fields}
\end{figure*}

\subsection{Sensor Placement and Particle Tracking}
\label{sec:sensor_placement}
To track the movement of motion particles we distribute a set of sensors~$S$ in the interaction space\changed{~$U$}\erase{of~$M$ and~$D$}. As we cannot assume any prior knowledge about where an interaction is going to happen (\ie~for simulations), we distribute sensors in the entire domain of $U$, by subdividing the space with an Octree. \changed{To emphasize the importance of motions in the vicinity of $M$, we produce a graded set of cubes that gets finer closer to the shape surface. As we cannot rely on the tessellation of $M$ to produce the levels of the Octree, we sample the surface with a 3D variant of \textit{Poisson Disk Sampling}. We randomly select a triangle $T$ of $M$ and compute a sample candidate by
\begin{equation}
e_c = (1-\sqrt{r_1})T_a + \sqrt{r1}(1-r_2)T_b + (\sqrt{r_1}r_2)T_c \,,
\end{equation}
where $T_a$, $T_b$, $T_c$ are the vertices of the triangle $T$ in ${\rm I\!R}^3$ and $r_1$, $r_2$ are random numbers in $[0,1]$~\cite{Osada:2002:SD:571647.571648}. We add $e_c$ to the sample set $E$ when $\forall e_i \in E: |e_i - e_c| > c$, where $c$ is a user defined threshold that specifies the minimum distance between the samples. We then insert the samples into the Octree and subdivide each of its cells as long as they contain more than one sample. Thereby, threshold $c$ is a control parameter for the depth of the Octree and hence implicitly defines the number of cells~(Figure~\ref{fig:sensor_placement}). For most of our tests we used sample distances from 0.01 to 0.5, which roughly corresponds to $1\%$ of the width of the interaction space $U$}.

\changed{Sensors are placed at the cell centers of the Octree, each observing a region defined by the corresponding cell.} This allows us to maintain cuboidal sensor regions, while the sensors are adaptively placed according to their distance to the input shape. \tog{Intuitively, motions close to the surface contribute the most in describing the interaction; however, we found that distant motions, \togerase{a head tilt to look down at a chair before sitting down on it,} can also be important for thoroughly describing an interaction. This can be seen in Figure~\ref{fig:agent_placement}, (a-c): the head tilt of the model produces unique motion trajectories (visualized in red) that help to discriminate sitting interactions from other motion types.} Moreover, the arrangement of shape parts of $M$ affects how the motion driver interacts with the shape --- some regions are more important than others. \changed{Employing an adaptive spatial subdivision allows us to place sensors with different size sensing regions and thereby to capture particle motions with varying degrees of precision.}\erase{We account for this by placing sensors with different size sensing regions and by capturing particle motions with varying degrees of precision.}


Each sensor $s_i$ defines a cuboidal region $r_{s_{i}}$ which tracks bypassing motion particles. We sample each particle's motion trajectory in $r_{s_{i}}$ with a constant time step $\Delta t$ and retain its position and velocity as it moves through the volume over time. \erase{After capturing the interaction of all particles, we}After capturing the interaction, we uniformly discretize $r_{s_i}$ \changed{into cells $c_{i,j}$} and average the positions and directions of all trajectory samples that fall into the same \erase{sensor volume}\changed{cell $c_{i,j}$ resulting in the vector $\vec{u_{c_{i,j}}}$}. For each \changed{sensor} $s_i$ we build a vector field $v_i$ through trilinear interpolation of the discretized sensor region. The set of vector fields $V=\{v_1, ... v_n\}$ represents the observed interaction of two participating shapes --- so we can now proceed to analyze the properties of the interaction by detecting patterns in the set of such vector fields.

\vspace{2mm}
\subsection{Encoding Motion Flows}
\label{sec:encoding_motion_flows}
\changed{Each interaction is represented as a set of vector fields.} To facilitate the comparison and clustering of interactions we\erase{must} define a \changed{distance measure}\erase{relative distance} \changed{between two sets of vector fields $V_1$ and $V_2$, each representing one interaction of a motion driver and a static object}. To compute \erase{the} \changed{this} distance we \erase{will} compare locally captured motion flows. We compute the gradient for each vector field $v_i \in V_1, V_2$ \erase{and} \changed{to} derive a set of its first-order attributes~\cite{Liefei2011}, \changed{which associate a real  value to each cell $c_{i,j}$}. In particular, \changed{we measure the magnitude of the attributes vorticity, dilatation, shear strain rate, and the gradient tensor field $T$ (see Appendix~\ref{app:attributes} for the precise definitions). Additionally, we consider the magnitude of vector $\vec{u_{c_{i,j}}}$ and the dot product of $\vec{u_{c_{i,j}}}$ with the closest surface normal of the mesh as two vector attributes. Computing this set of attributes on $v_i$ allows us to quantify the measured interactions $I_1$ and $I_2$ by computing a set of histograms.}

\begin{figure*}[t]
\includegraphics[width=\linewidth]{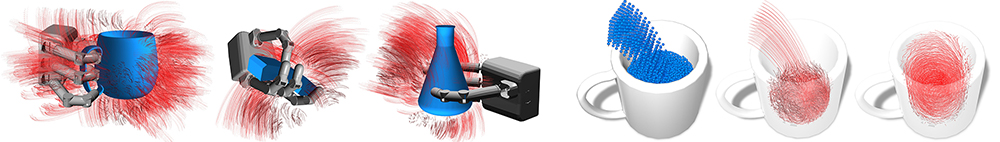}\hfill
\vspace{-4mm}
\hspace*{18mm}(a)\hspace*{33mm}(b)\hspace*{11mm}(c)\hspace*{29mm}(d)\hspace*{24mm}(e)\hspace*{23mm}(f)\hfill~
\caption{The simulation of a humanoid hand grasping small scale man-made objects: cup (a), phone (b), and bottle (c). The motion trajectories indicate a series of interactions performed on the corresponding object. Interactions of a fluid and a cup: the SPH fluid particles are poured into a cup (d), and their trajectories are captured (e). The trajectories of a stirring motion from a different interaction~(f). }
\vspace{-2mm}
\label{fig:graspit_fluid}
\end{figure*}

\textbf{Histograms.} We compute a histogram for each attribute $a$ of the vector field $v_i$, forming a set of histograms
\begin{equation}
{\cal H} = \{h_a, a \in A\},
\end{equation}
where \(h_a\) is a normalized histogram over the attribute $a$ (Figure~\ref{fig:flow_fields},~a).
The histogram $h_a$ is built by binning the corresponding attribute values of a vector field in the range~$[0,1]$. \changed{The contribution to the individual bins is weighted by an exponential scaling of the Euclidean distance of the relative position $p_{c_{i,j}}$ of cell $c_{i,j}$ to the closest point $p_M$ on the surface of $M$. We compute the value of a given bin $m_k$ by
\begin{equation}
m_k = \frac{1}{N}\sum_{j=1}^{n_c} t_j\exp(-(\frac{\|p_{c_{i,j}} - p_M\|}{r})^2) \cdot \chi_{E}(t_j),
\end{equation}
where $n_c$ is the number of cells in the vector field, $t_j$ the attribute value in the current cell, $r$ the size of the sensor~$s_i$ and $\chi_E$ the indicator function that associates $\chi_E=1$ if $t_j\in E$ and $\chi_E=0$ otherwise, where $E=[\frac{k}{N}, \frac{k+1}{N}[$.} Each bin is normalized by $N$, the number of all bins. The exponential weighting ensures that distant motions have less influence than those in the vicinity of $M$. \changed{Finally, we compute a set of global histograms
\begin{equation}
{\cal \overline H} = \{\overline h_a, a \in A\},
\end{equation}
where each histogram~$\overline h_a$ represents the average of all histograms~$h_a$ of the attribute $a$ across all vector fields $v_i \in V$ (Figure~\ref{fig:flow_fields},~b). To account for differently sized sensor regions we compute each bin $\overline m_{k}$ of $\overline h_a$ as
\begin{equation}
\overline m_{k} = \frac{\sum_{j=1}^{n} m_{j_k} r_j^3}{\sum_{j=1}^{n} r_j^3},
\end{equation}
where $n$ is the number of histograms (and vector fields) of one attribute, $m_{j_k}$ the $k$-th bin of the $j$-th histogram $h_a$ and $r_j$ the size of the corresponding sensor $s$. Additionally, we only consider sensors that tracked motion particles to reduce the complexity. \tog{For the sake of a clear presentation we describe the computation of the histograms as a two stage process, however, the global histograms can be directly computed from the vector fields.}

 \hide{of the computation and to maintain significant features in the histograms.}} \erase{This reduces the complexity of computation and lowers the noise introduced through less meaningful motions. }

\changed{\textbf{Interaction Descriptor.} We define our interaction descriptor as a set of global histograms, encoding the motion patterns of an interaction of a motion driver and a static object. We compare interactions (Figure~\ref{fig:flow_fields},~c) by computing the relative distance between two sets of histograms as the similarity score:}
\erase{\textbf{Motion signatures.} To compare two interactions $I_1$ and $I_2$ we calculate a similarity score }
\begin{equation}
d(V_1,V_2) = \frac{\sum_{a \in A}w_{a} D_B(h_{a},k_{a})}{|A|},
\end{equation}
where \(w_{a}\) is a weight for attribute \({a}\), $|A|$ the number of attributes and \({D_B}\) the Bhattacharyya distance between two global histograms \(h_{a}\) and \(k_{a}\), \erase{that are the averages of all histograms} associated with \(V_1\) and \(V_2\) respectively. \changed{Operating on the global histograms allows us to compare interactions, even if the number of used sensors for each of them differs.} \erase{The similarity score $d$ is the relative distance between two interactions; it is non-negative and symmetric in its arguments.}We use the Bhattacharyya distance, as it has proved effective at relating semantically similar interactions and discriminating ones that differ. Moreover, we found the weights \(w_{a}\) by experimentation; dilatation magnitude and vector orientation turn out to be the most important attributes, while shear strain rate and vector magnitude do not contribute much to interaction comparison.

\textbf{Time variance.}\label{sec:time_var} \erase{We need to} \changed{A time sensitive descriptor allows to differentiate interactions of varying speeds and time-steps, an important property for dynamic interactions. However, while exploiting this features is desirable in many scenarios, we aim at providing a means of controlling the importance of temporal information. For example, when grasping a cup, the speed of the grasp might differentiate motions, while it can be irrelevant for classifying the shape. Our approach can accommodate both modalities. When discretizing the sensor boxes into vector fields we compute a vector $\vec{u_{c_{i,j}}}$ for each cell $c_{i,j}$ in the vector field $v_i$. This vector is the sum of all trajectory vectors located in $c_{i,j}$ normalized by the factor~\(F\):}
\begin{equation}
\vec{u_{c_{i,j}}} = \frac{\sum_{u_k \in c_{i,j}}u_k}{F}.
\end{equation}
\changed{To differentiate similar interactions with different speeds, we choose $F = |c_{i,j}|$, effectively calculating the average vector, which retains the velocity of the motion flow. When we match similar interactions independent of their temporal resolution we choose \(F= \|\sum_{u_k \in c_{i,j} }u_k\|\), which normalizes the vectors and thus only maintains information of the motion flow direction. }



\textbf{Alternative Approaches.} Instead of averaging the captured \erase{distances}\changed{motions} we performed a number of experiments with more sophisticated mathematical methods for finding matching pairs of vector fields, \eg through \textit{Hungarian Optimization}\hide{and \textit{Spectral Alignment}~\cite{1544893}} and tested different distance measures such as Chi-Square or Earth-Mover-Distance~\cite{710701}, but found averaging the signatures performs better on our data sets. Moreover, we employed \textit{Dynamic Time Warping (DTW)}~\cite{Black97automaticallyclustering} to temporally align the histograms before computing the signatures, however, we found that in most cases DTW did not improve much the quality of the clustering.

\begin{figure*}[t]
\includegraphics[width=\linewidth]{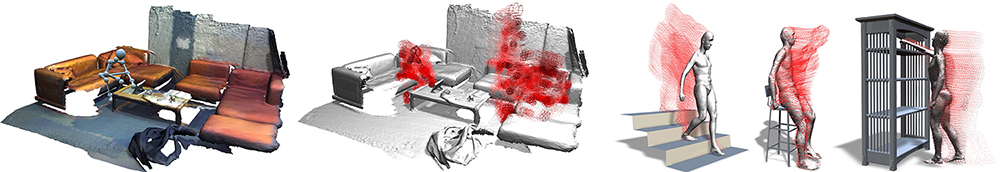}\hfill
\vspace{-4mm}
\hspace*{8mm}(a)\hspace*{50mm}(b)\hspace*{47mm}(c)\hspace*{16mm}(d)\hspace*{13mm}(e)\hfill~
\caption{Real world interactions: we used the SceneGrok [Savva et al. 2014] dataset which includes 45 RGB-D captures in 14 environments~(a) to capture motion trajectories of humans~(b) performing interactions that can be categorized into 8 classes. (c-e): motion capture data of a human interaction with physical objects can also be used in our framework.}
\label{fig:rgbd_mocap}
\end{figure*}

\section{Implementation and Datasets}

We implemented our method in C++ with OpenGL and GLSL. The system was deployed on a desktop computer with an Intel Xeon processor clocked at 3.7~GHz, with 32~GB~RAM and an NVIDIA GTX~980~GPU. Results  shown in the paper were generated with our own framework.

We focus on capturing time-variant interactions of one or multiple motion drivers and a single observed object. The motion trajectories are generated by tracking the motion driver (surface or particles) and by updating the corresponding motion particles efficiently through graphics hardware. Depending on the number of motion particles on the surface and the amount of tracked trajectories, this can be done at interactive rates \tog{\togerase{or even in real-time}}. We exploit the interactivity of our system to provide the user with control of capturing  the motion trajectories. We provide access to the control parameters, such as the number of motion particles, sampling speed of simulations and animations, and parameters for the fluid simulations, \eg velocity, viscosity, and surface tension.

\begin{table}[t!]
\scriptsize
\begin{center}
\tabcolsep3pt
 \scalebox{0.99}{
    \begin{tabular}{c|c|c|c|c|c|c|c|c|c|c|c}
        {} & \multicolumn{3}{c|}{Motion Driver} & \multicolumn{2}{c|}{Object} & \multicolumn{3}{c|}{Sensors} & \multicolumn{2}{c|}{Trajectories} & \multicolumn{1}{c}{Signature}\\
        \hline
        {\bf Fig.}                        & {\bf MP} & {\bf MT} & {\bf MY} & {\bf OS}  & {\bf OT} & {\bf S} & {\bf OL} & {\bf DS} & {\bf T} & {\bf TS} & {\bf ST}\\
        \hline
        \ref{fig:agent_placement} (a)        & 2.0k  & 552s  & AN  & 0.7k  & 0.40s      & 2.6k  & 8    & 6   & 17k  & 991k   & 0.54s \\ 
        \ref{fig:agent_placement} (b)        & 2.0k  & 552s  & AN  & 0.3k  & 0.20s      & 1.6k  & 8    & 6   & 13k  & 896k   & 0.39s\\ 
        \ref{fig:shape_retrieval} (*)        & 2.0k  & 581s  & AN  & 1.1k  & 0.43s      & 3.3k  & 8    & 6   & 14k  & 907k   & 0.65s\\ 

        \ref{fig:graspit_fluid}   (e)        & 2.0k  & -     & FL  & 5.5k & 2.23s     & 17.7k & 7    & 5  & 28k  & 601k   & 2.82s\\ 
        \ref{fig:human_fluid_mds} (*)        & 2.0k  & -     & FL  & 3.6k  & 1.76s     & 4.5k  & 8   & 10  & 10k  & 242k    & 1.74s\\ 
        \ref{fig:shape_retrieval} (\dag)     & 2.0k  & -     & FL  & 1.8k   & 0.65s   & 4.8k  & 6    & 6  & 20k  & 570k   & 0.65s\\ 
        \ref{fig:saliency_estimation} (*)    & 2.0k  & -     & FL  & 0.8k   & 0.21s   & 2.9k  & 8    & 10  & 22k  & 566k   & 0.57s\\ 

        \ref{fig:graspit_fluid} (a)          & 1.5k  & 161s  & GR  & 0.9k   & 0.56s    & 3.2k  & 7    & 5   & 18k  & 201k  & 0.81s\\ 
        \ref{fig:human_fluid_mds} (\dag)     & 1.6k  & 154s  & GR  & 1.6k   & 0.75s    & 5.6k  & 8    & 5   & 21k  & 294k  & 1.63s\\ 

        \ref{fig:rgbd_mocap} (a)             & 1.8k  & 80s   & SG  & 1.8k   & 1.75s    & 6.6k  & 10   & 15  & 51k  & 639k   & 1.11s\\ 
        \ref{fig:rgbd_mocap} (d)             & 2.0k  & 119s  & MC  & 0.5k   & 0.18s    & 2.1k  & 9    & 8   & 5k   &  79k   & 0.34s\\ 
        \hline
    \end{tabular}
    }
\end{center}
\caption{
Details on the processed interactions of some of the figures shown in the paper; MP=Motion Particles, MT=Motion Driver Sample Time, OS=Object Samples, OT=Object Sampling Time, S=Sensors, OL=Octree Level, DS=Domain Size, T=Trajectories, TS=Trajectory Samples, ST=Signature Time. The table is sorted by the type (MY) of motion driver: AN=Animation, FL=Fluid Simulation, GR=GraspIt! Simulation, SG=SceneGrok, MC=MotionCapture. For all our tests we sampled motion trajectories with $\Delta t = 25ms$ and set the sample distance for the observed object to 0.1. }
\label{table:statistics}
\vspace{-4mm}
\end{table}

\tog{The initialization step is done during preprocessing and the most time demanding step is the farthest point sampling strategy of the motion driver that is currently implemented without any acceleration and has $O(n^2)$ complexity, where $n$ is the number of particles. The most demanding part of the runtime module is the tracking of motion particle trajectories. We test if a motion particle is contained in a sensor volume and capture its motion trajectory, resulting in at worst $O(mn)$ complexity, where $m$ and $n$ is the number of sensors and particles, respectively.}

\changed{We measured the sampling of motion drivers with 80--600 seconds (with 2k particles as the upper limit) and 2--20s for observed objects. Depending on the recording time, we capture around 100k--8M trajectory samples which roughly amounts to 10--80MB of motion data.} \tog{Computing the signatures, which involves calculating vector fields and their corresponding histograms from the trajectories takes 0.30--3.0 seconds.}\erase{Comparing two signatures to compute their relative distance (Section~\ref{sec:encoding_motion_flows}), takes less than 1~ms}~Table~\ref{table:statistics} provides details on the complexity of some of the interactions shown in the paper.  


\subsection{Datasets}
Shapes can interact in a variety of ways, which makes comparing and aggregating them a challenging task. \changed{To show the versatility of our system we employed data from various sources. Simulations and RGB-D data provide more meaningful signatures compared to precomputed data sets.} This is mostly because rigged animations and motion capture data cannot be easily acquired, resulting in a sparse representation of an interaction when used for computing the distances. Simulations can be used with arbitrary input meshes and can dynamically adapt. Moreover, we found that aggregating repeated interactions of the same two participating shapes improves the quality of the resulting signatures.

\textbf{Simulations.} We explored two different types of simulation.
First, we explored interactions of a human hand grasping objects. We used the \textit{GraspIt!} simulator~\shortcite{graspit:2012}, which is a publicly available tool for simulating grasping motions of human and robot hand designs. It accommodates arbitrary input meshes and detects the quality of grasping configurations by simulating collisions of hands and obstacles. We used \textit{GraspIt!} to simulate the interaction of human and robotic hands with small scale man-made objects, \eg cups, cellular phones, and books (Figure~\ref{fig:graspit_fluid}, a-c). \changed{We do not simulate any advanced effects, such as force closure or precision grasps, but instead rely on \textit{GraspIt!'s} support for detecting collisions and determining contact.}

\begin{figure}[t]
\begin{centering}
\includegraphics[width=\columnwidth]{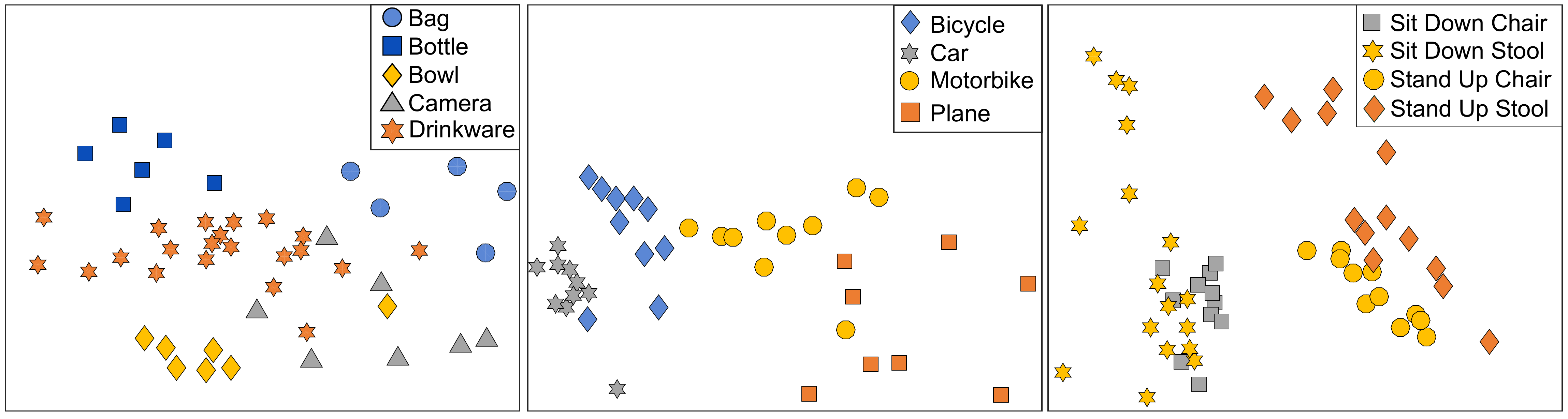}\hfill
\vspace{3mm}
\includegraphics[width=\columnwidth]{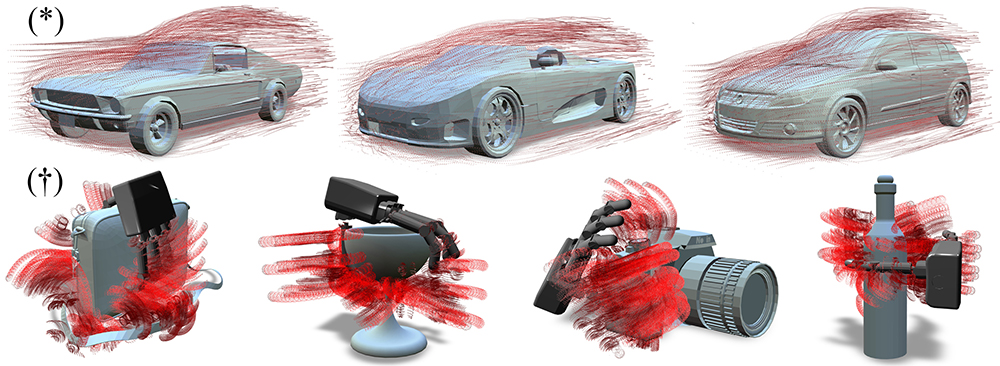}\hfill
\caption{Top row: multidimensional scaling (MDS) plots of isolated clusters of categories. \changed{Left: interactions of simulated grasps from \textit{Graspit!} and small scale objects. Middle: a plot of interactions of objects in wind tunnels. Right: motion captures of a human sitting on furniture. Our method generates discernable clusters, objects of overlapping classes share functional commonalities, also noticeable from their visual appearance. Middle + bottom:  objects and their motion trajectories used for the MDS embeddings.}}
\label{fig:human_fluid_mds}
\end{centering}
\end{figure}

Second, we simulated fluids with Smoothed Particle Hydrodynamics (SPH)~\cite{monaghan1992smoothed}.\erase{Unlike Eulerian approaches, particles are trackable in 3D space and thus easily allow detecting fluid-obstacle collisions.} We randomly select fluid particles to capture motion trajectories of the fluid flow. We used SPH to explore objects interacting with fluids, such as a fluid poured in a cup, and wind fields, such as planes and cars. \changed{Figure~\ref{fig:graspit_fluid}~(d-f) and \ref{fig:human_fluid_mds} (middle row) show examples of fluid flows and the captured motion trajectories.}

\textbf{Animations.} To account for full-body interactions of humans with large scale rigid objects such as furniture, we used animated humanoid meshes and motion captured data. We employed models from Adobe's Mixamo~\shortcite{mixamo:2015} model repository that provides a large number of humanoid models rigged to perform certain actions, \eg jumping, sitting on a chair, or walking. We manually fitted these models to object meshes and captured their motion data when performing the predefined action. Figure~\ref{fig:human_walks} shows three human models performing different motions.

\textbf{Motion Capture and RGB-D.}
\changed{To show that our method works with real-world data, we employed the RGB-D scans from SceneGrok~\cite{Savva:2014:SIA:2661229.2661230} and motion capture (MoCap) data from the KIT Whole-Body Human Motion Database~\shortcite{HumanMotionDB:2015}, illustrated in Figure~\ref{fig:rgbd_mocap}. The SceneGrok dataset contains 45 RGB-D captures in 14 environments representing humans performing interactions that can be categorized into 8 different classes. We use the captured environment as static object and build a humanoid mesh that is updated based on the provided skeleton positions. The produced mesh is used as the motion driver.

The MoCap dataset includes high-quality human whole-body motions that involve human-object interactions. In particular, we explored 10 sitting interactions each with 3-4 objects. To generate a surface model we fit the human body model (template mesh)~\cite{Anguelov:SIGGRAPH05} to MoCap data points. Specifically, we embedded a skeleton to the template using automatic rigging software~\cite{Baran:2007:ARA} and scaled the body segments of the template to fit the body shape of the captured subject. For both data types we sample point locations from the produced surface mesh (Section~\ref{sec:motion_driver}) which is continuously updated to drive motion to capture interaction trajectories.
}


\section{Results, Evaluation and Applications}
\label{sec:eval}

We have introduced a novel descriptor for aggregating time-variant interactions of arbitrary shapes and motion drivers for 3-dimensional spaces. \changed{Unlike previous approaches we are not limited to static scenes and rigid surfaces but instead allow and exploit time variance for interactions}. In this section we show results, provide an evaluation of our method and discuss its performance against other state-of-the-art shape and interaction descriptors. Moreover, we show that knowledge of interactions informs knowledge of the shape function that can be employed for shape classification and saliency estimation.

\subsection{Results}
Figure~\ref{fig:human_fluid_mds} (top row) shows two \textit{multidimensional scaling (MDS)} embeddings of simulated interactions.
\changed{We see a plot of grasping interactions from \textit{Graspit!} (left), objects in wind tunnels (middle), and a clustering of motion capture interactions (right). For both experiments our descriptor provides a meaningful aggregation of the interaction of shapes that allows us to relate and even retrieve them according to their function (Figure~\ref{fig:shape_retrieval}). For the wind experiment, bicycles and motorbikes share some overlap due to their similarities in both their function and shape. }




\begin{figure}[t]
\includegraphics[width=0.52\columnwidth]{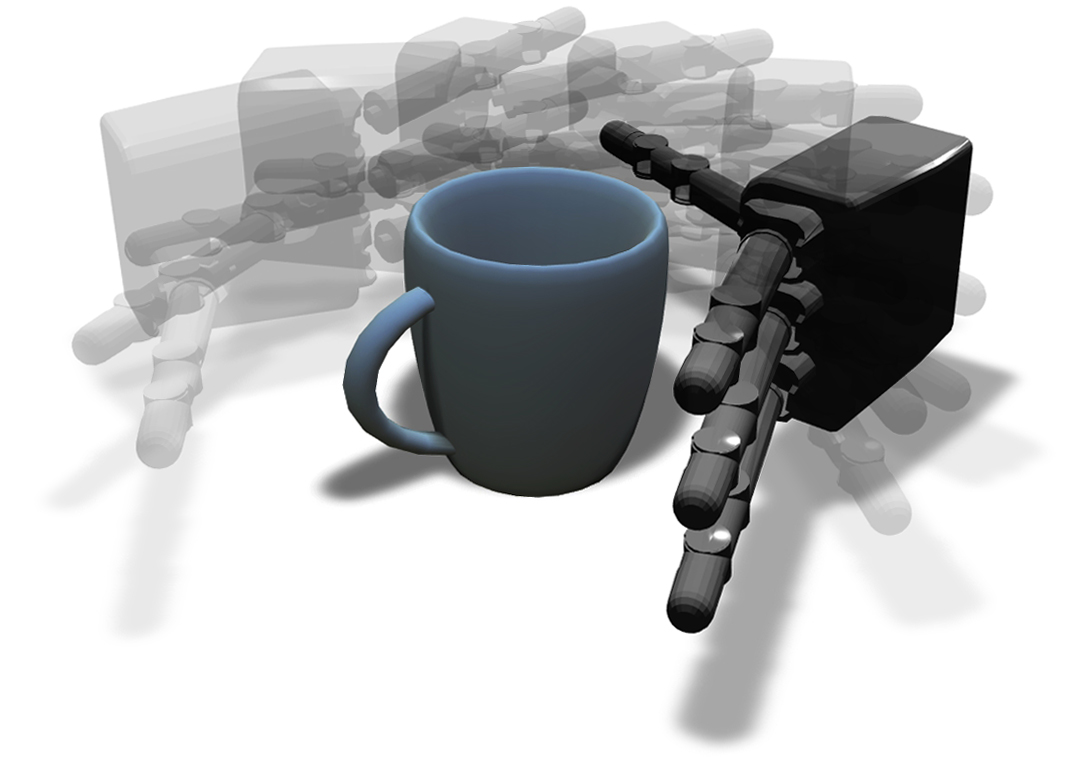}\hfill
\includegraphics[width=0.48\columnwidth]{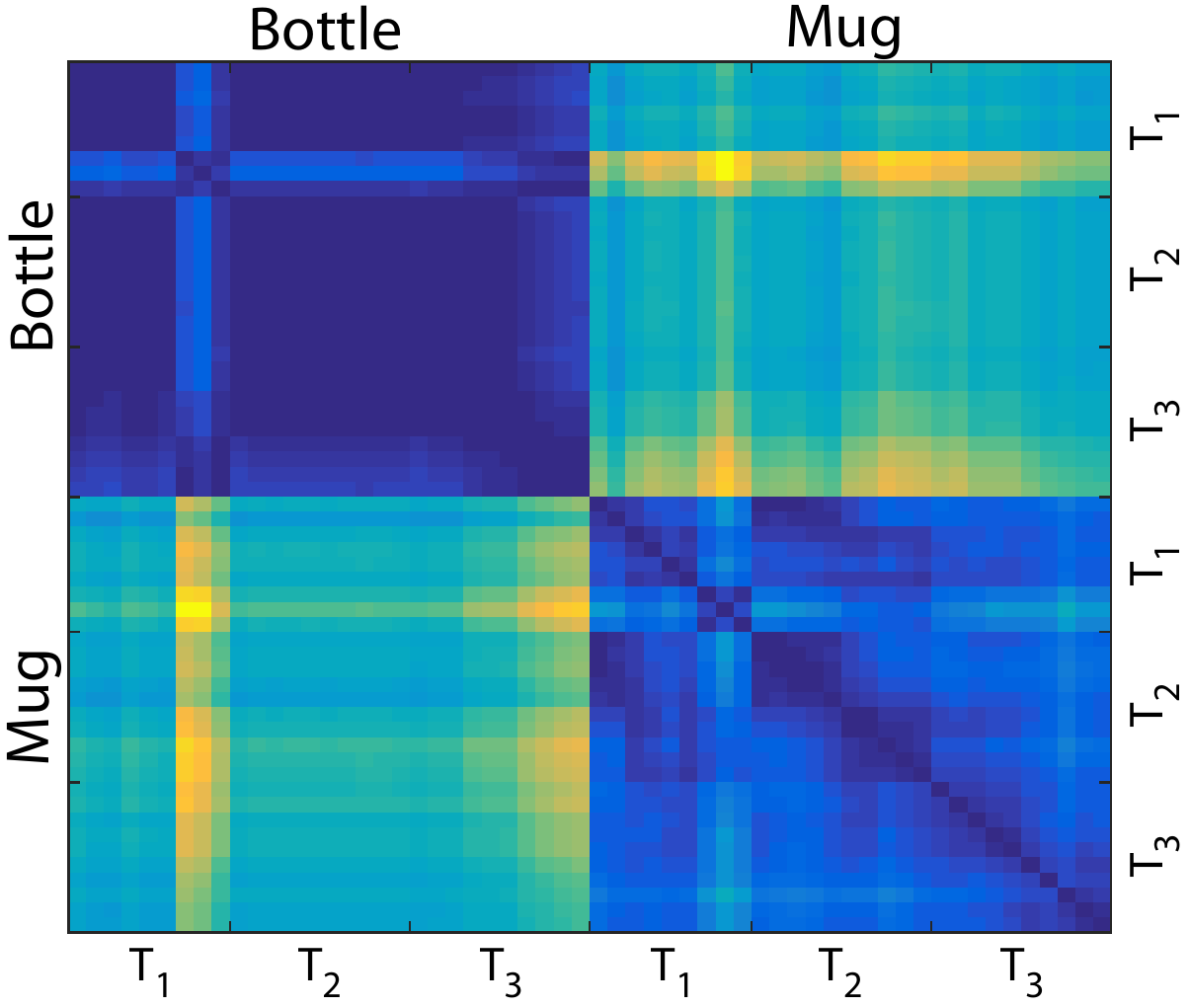}\hfill
\caption{Left: a subset of orientations of the motion driver around the up-axis of the observed shape. Right: a color-coded matrix that visualizes the distances between each of the configurations. We performed three experiments: rotation of the object around the up-axis ($T_1$), both of the other two axes ($T_2$) and the rotation of the motion driver around the object ($T_3$).}
\label{fig:rotation_behavior}
\vspace{4mm}
\includegraphics[width=0.43\columnwidth]{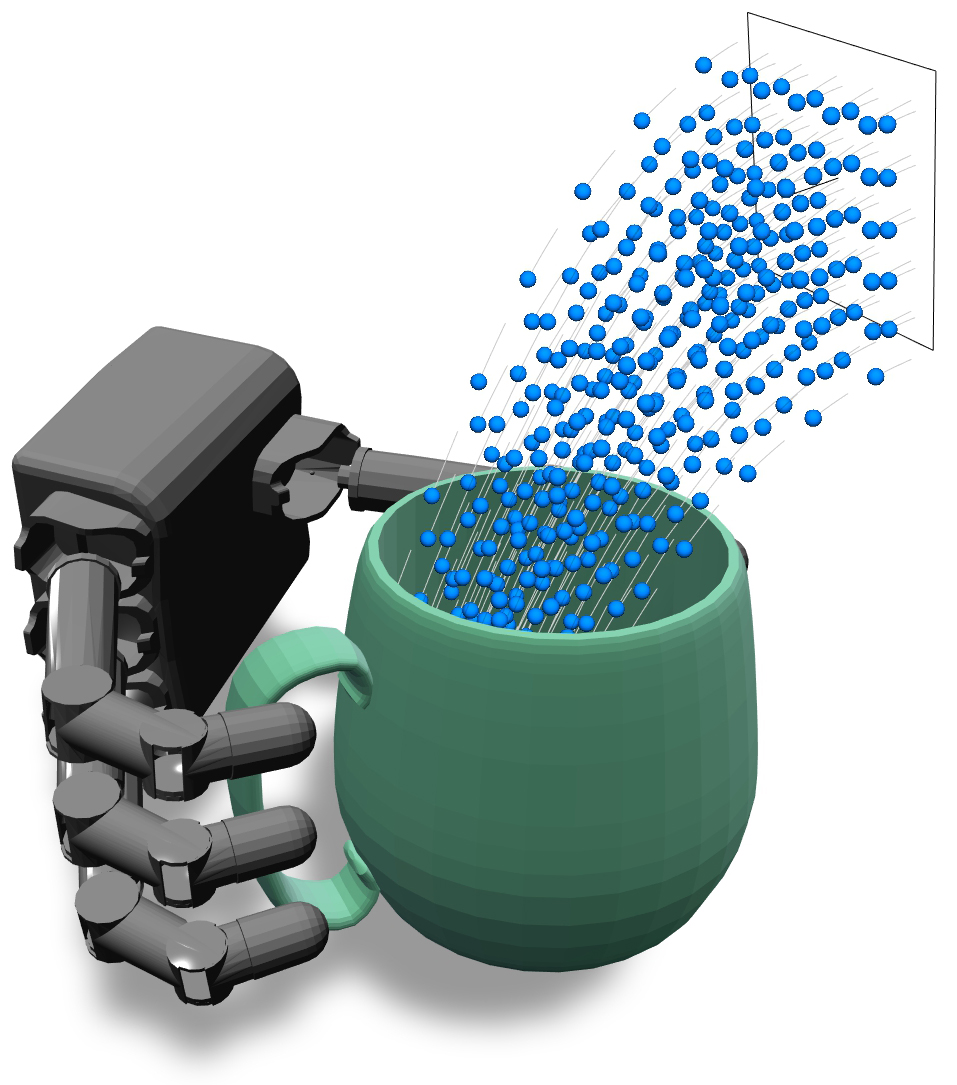}\hfill
\includegraphics[width=0.5\columnwidth]{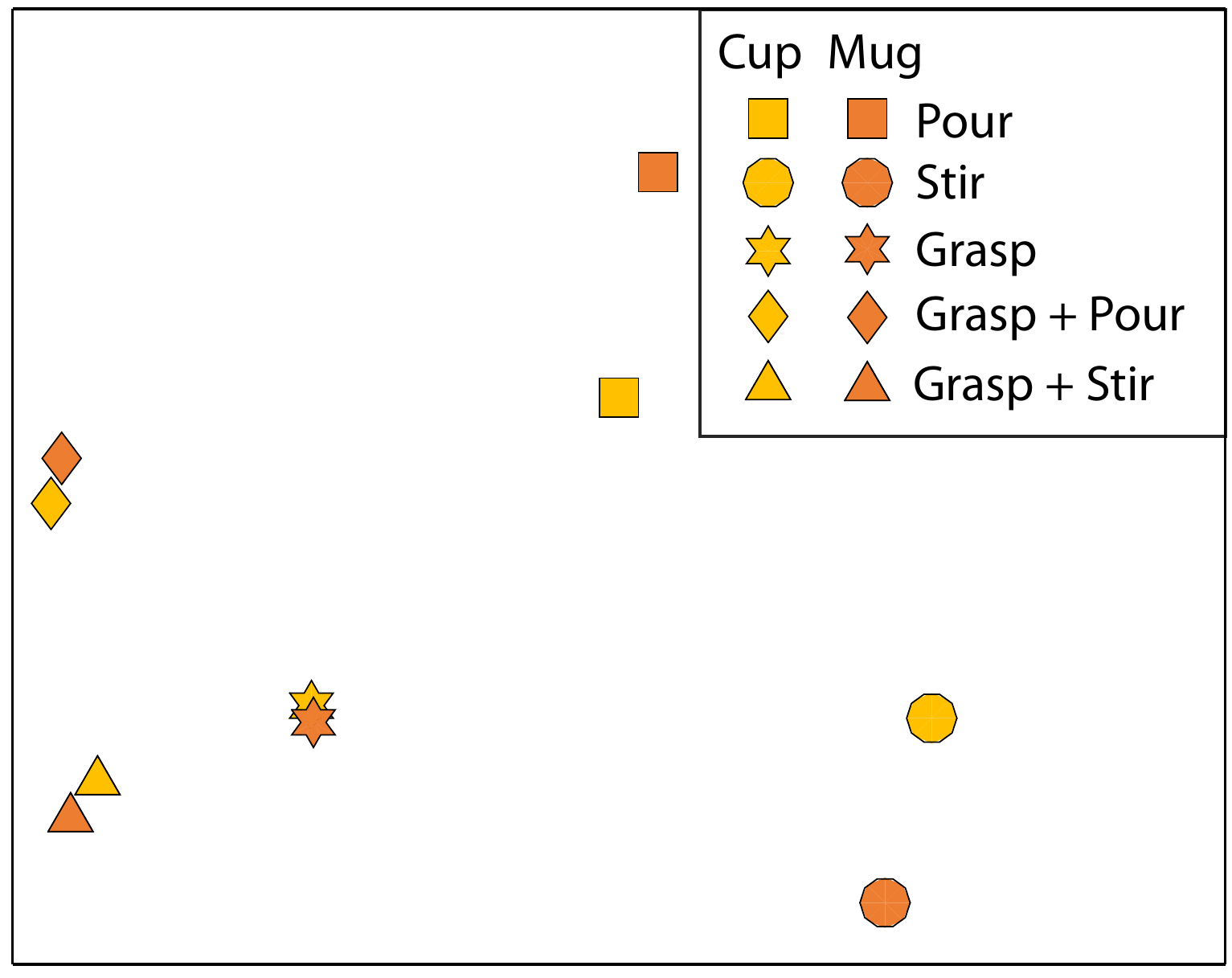}\hfill
\caption{The result of encoding two motion drivers. Left: the hand and the fluid are interacting with the observed cup at the same time. Right: a MDS embedding of multiple scenes showing the effect of using two motion drivers.}
\label{fig:multiple_motions}
\end{figure}

For most of our tests we randomly selected models from the Modelnet~\cite{DBLP:journals/corr/WuSKTX14} and ShapeNet~\cite{2015arXiv151203012C} model repositories\changed{; our database includes 157 objects of 20 classes.} We tested our framework with an automated setup that loads scene files with the specification of motion driver and observed object. We run the interactions for around 60 seconds to capture enough trajectories; animations and \textit{Graspit!} sets are tracked for their entire duration. \erase{We randomly selected the position and size of the emitter for \textit{swirling} and \textit{pouring} fluid simulations, but so that the fluid is poured into the cup.} To capture the interaction \erase{with} \changed{of} cars, planes and motorbikes with wind fields we setup wind tunnel scenes with fixed particle emitters.\erase{which explains the strong grouping of planes, cars and motorbikes in Figure 10.} For \textit{Graspit!} we randomly chose the positioning of the hand relative to the observed shape for the simulation and aggregated the interaction into a data set readable by our framework.

\tog{Figure~\ref{fig:rotation_behavior} illustrates the behavior of our descriptor with respect to variations in the orientation of observed object and motion driver. On the left we show a subset of arrangements of a hand positioned around the up-axis of the cup. The color-coded distance matrix~(right) visualizes the distances of all scenes we used in this experiment. We tested 58 (29 bottle, 29 mug) scenes representing different orientations of the observed object along the up-axis~($T_1$), both of the other axes~($T_2$), and of the motion driver~($T_3$). As we use global histograms to discriminate differences in the interactions, our descriptor is mostly agnostic to variations in the arrangements of shapes and involved motions.} 

\tog{Figure~\ref{fig:multiple_motions} shows a result of ten scenes of one and two motion drivers interacting with a single static object. Our method allows to capture several interactions at the same time and to encode them as a meaningful signature. The image on the left shows the setup for this experiment. A grasping motion and a fluid are both interacting with a cup as the observed model. The MDS embedding on the right shows that our method allows to produce a more principled clustering of the interactions when two motion drivers are used.}

\begin{figure}[t]
\includegraphics[width=0.48\columnwidth]{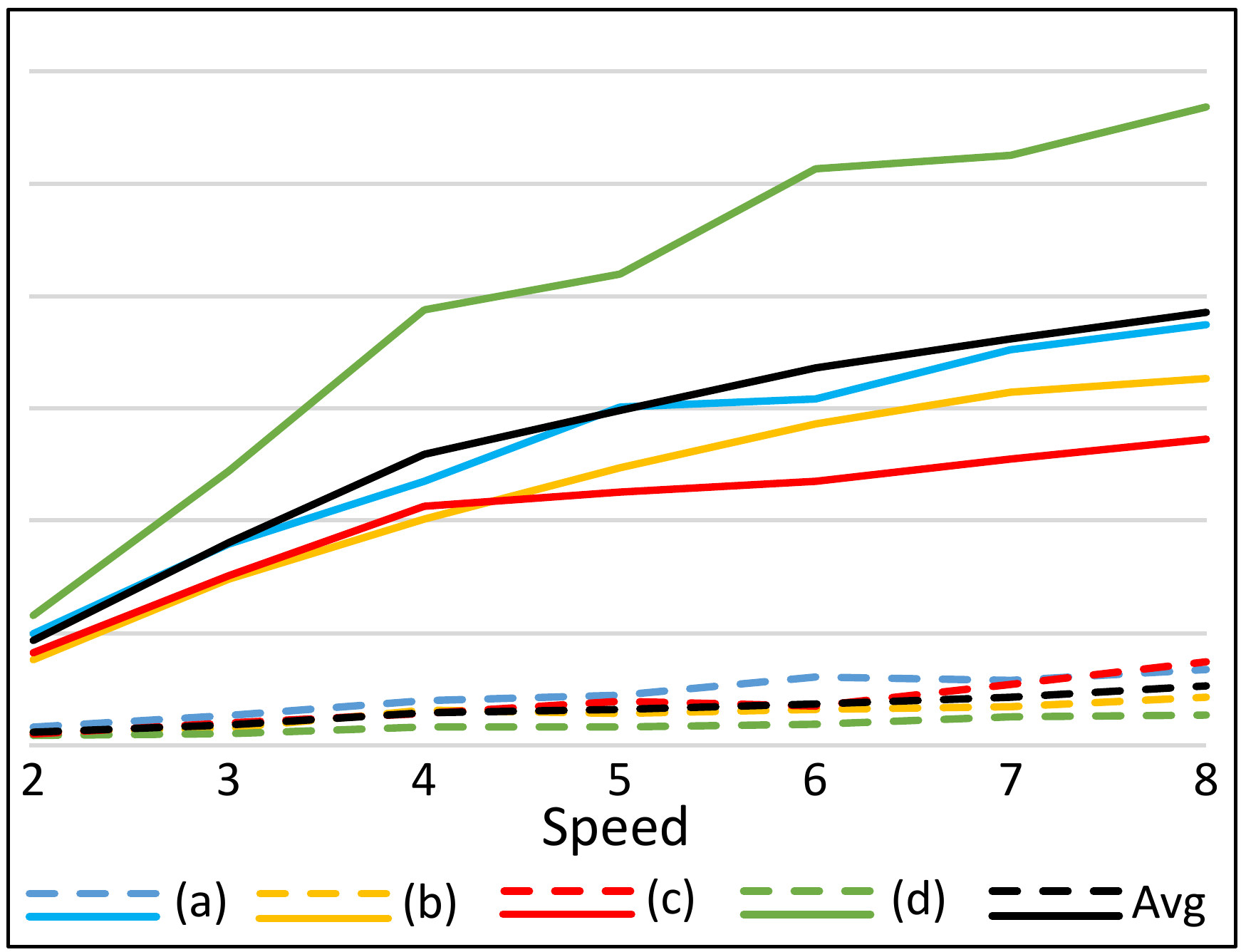}\hfill
\includegraphics[width=0.48\columnwidth]{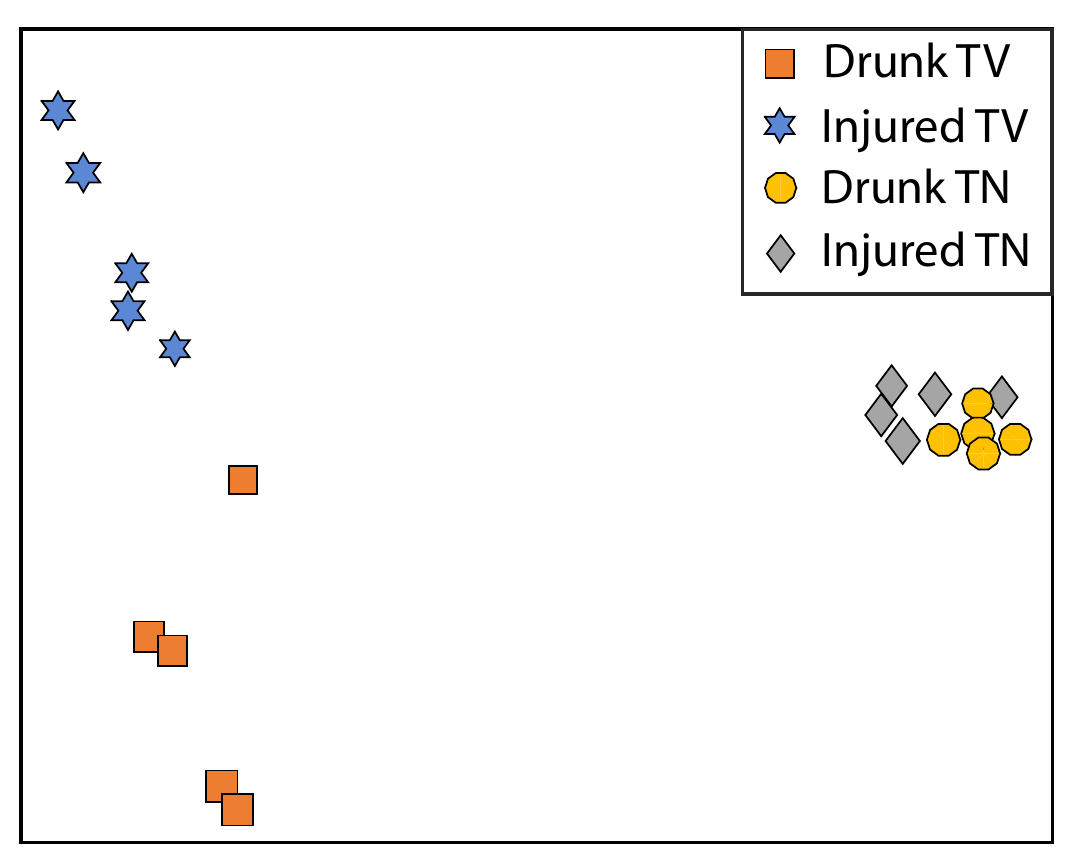}\hfill
\vspace{2mm}
\includegraphics[width=0.98\columnwidth]{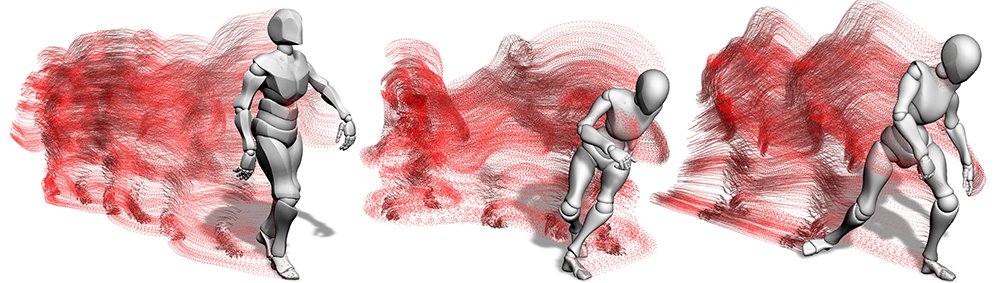}\hfill
\caption{Top left: relative distances of human-chair interactions (Figure~\ref{fig:agent_placement}~(a-d)) showing the effect of normalizing (dashed lines) or emphasizing (solid lines) the velocity of an interaction. Taking the speed into consideration allows us to even differentiate subtle motions. The MDS embedding (top right) shows 20  animations of "drunk" and "injured" walks from the Mixamo model repository (bottom) where we computed the distances with (TV) and without (TN) time variance. }
\label{fig:human_walks}
\vspace{-3mm}
\end{figure}

\changed{Finally, our descriptor enables to differentiate interactions based on their speed. Figure~\ref{fig:human_walks} (top, left) shows the relative distances of human-chair interactions shown in Figure~\ref{fig:agent_placement} (a-d) with different speeds. As can be seen in the diagram normalizing or emphasizing the velocity in the computation yields different results. Taking the velocities into consideration acts like a switch that allows to further differentiate subtle variations in the motion flows. The MDS embedding (top right) shows 20 animations of "drunk" and "injured" walks from the Mixamo model repository (bottom) where we computed the distances with (TV) and without (TN) time variance (Section~\ref{sec:time_var}). Taking velocities into consideration yields finer clusters while normalizing the motions causes a stronger grouping. \hide{For this test we used a static grid instead of an Octree for placing the sensors.}}

\subsection{Evaluation}
\erase{It is important to note that, u}Unlike previous methods, our descriptor uses time-varying data. Therefore, comparing it to existing techniques only has limited meaning. However, we measure its effectiveness for common shape classification and retrieval tasks. For these evaluations we employ precision and recall queries on our data set and compare the performance to existing descriptors on the same task\changed{, but with static scenes}. Additionally, we discuss a number of experiments to show the effect of parameters, \eg the number of motion particles and sensors, specified in our framework.

\textbf{Parameters and Method.} For most of our tests we distributed between a few hundred and a couple of thousand motion particles on moving surfaces or advect them directly through the simulation. We tested different settings for the number of motion particles and sensors and found that using only a few particles generates sparse motion trajectories that do not provide enough data to extract meaningful signatures. On the other hand, placing many particles does not improve the results because of the discretization of sensor regions into vector fields\erase{; speed and directions sampled in similar locations are averaged when processing the signatures}. We computed vector fields of sizes $4^3$, $8^3$, and $16^3$ and found that the method is somehow independent to the resolution in this range, however, size $8^3$ performs best.

\begin{figure}[t]
\includegraphics[width=\linewidth]{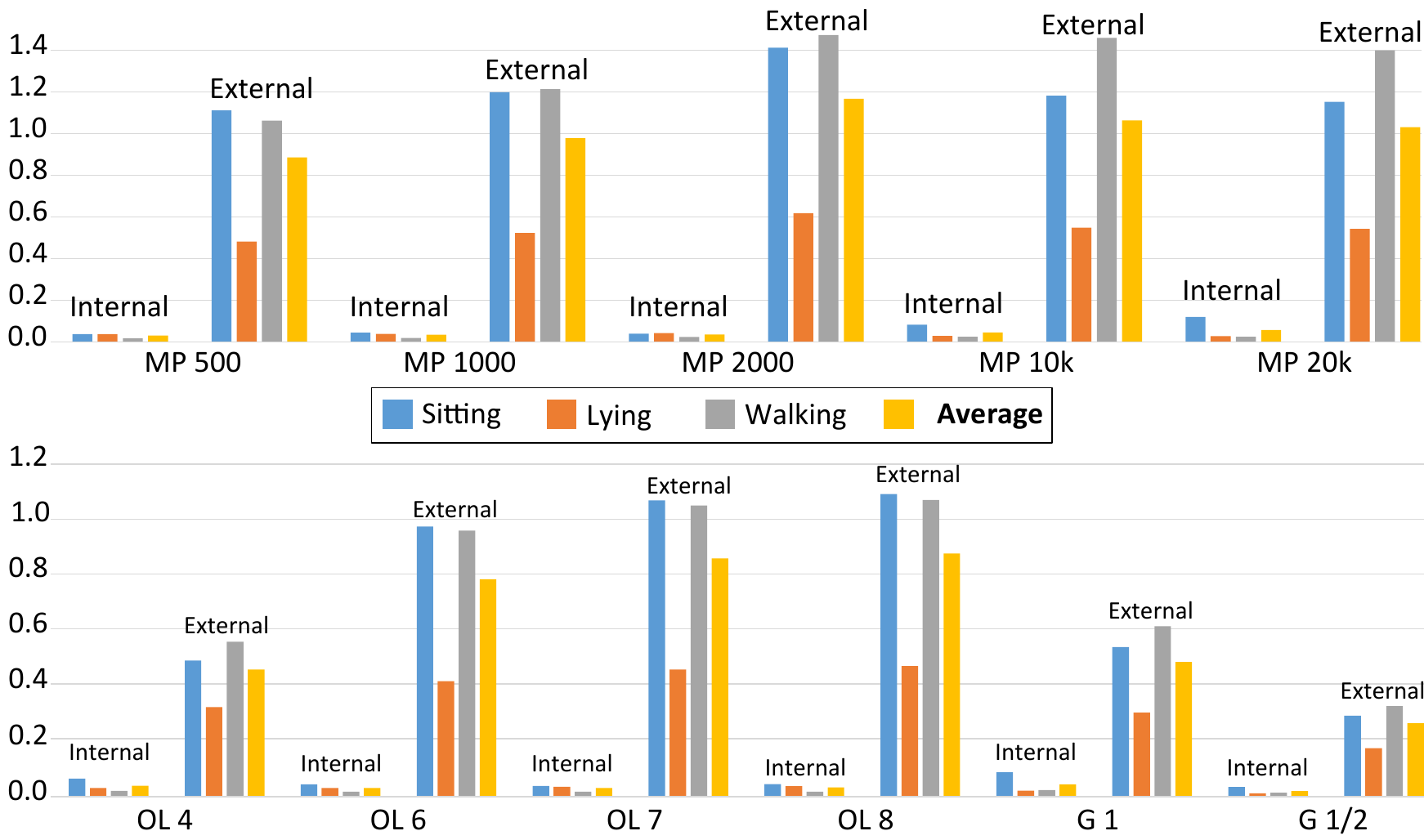}\hfill
\caption{Comparison of different settings for the number of motion particles and sensors in our framework. Top: the number of motion particles (MP); bottom: the number of sensors defined by the levels of an Octree (OL) and by a uniform grid (G). The diagrams show the relative distances of interactions within the same group (internal) and across different groups (external). We used five interactions per category, where we used different animations as motion driver for the interaction with chairs and beds (sitting, lying) or the floor (walking).}
\label{fig:comparison_bars}
\vspace{6mm}
\centering
\includegraphics[width=0.5\columnwidth]{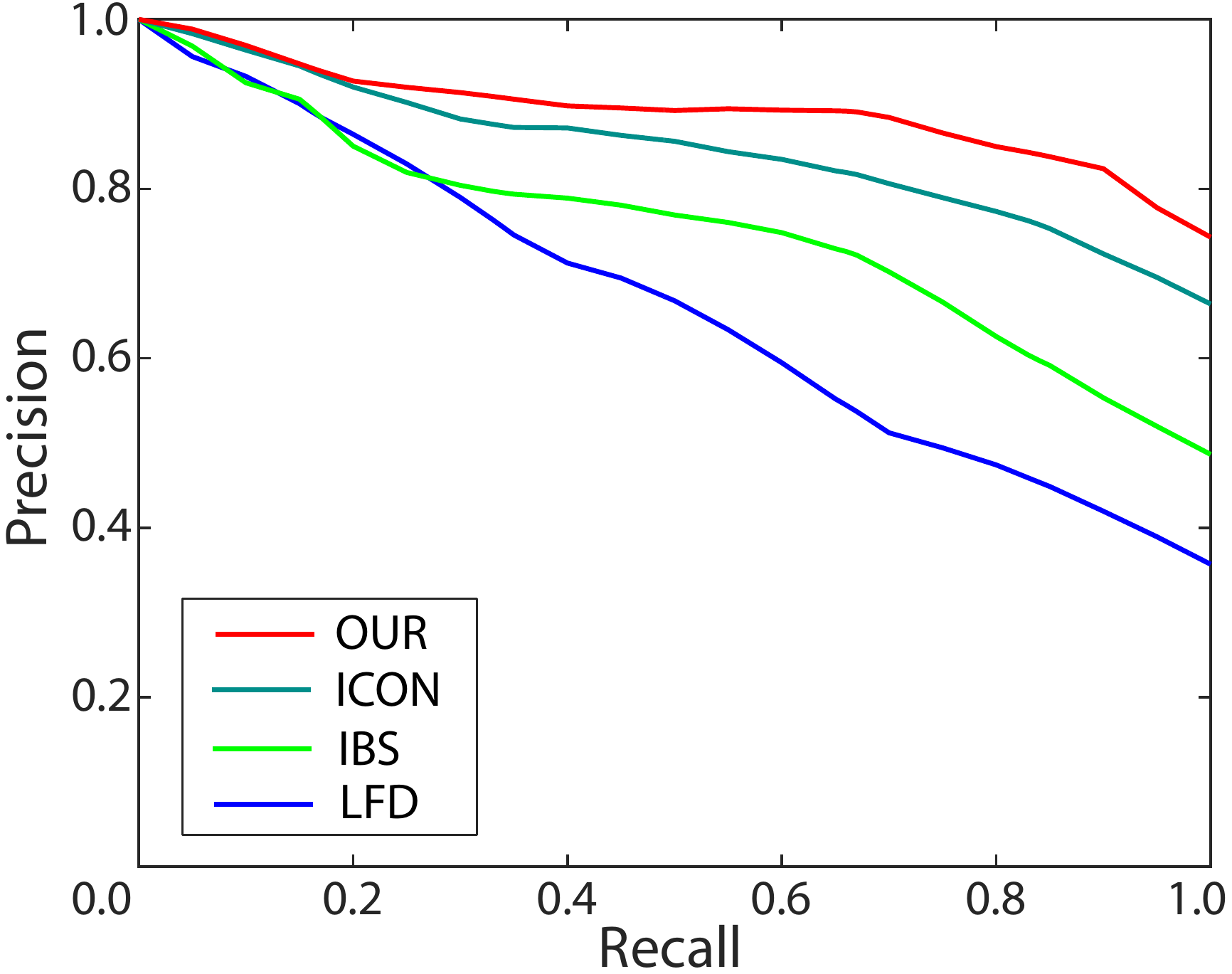}\hfill
\caption{Precision and recall diagram showing the performance of our descriptor compared to the LFD, IBS and ICON.}
\label{fig:pr_diagrams}
\vspace{-4mm}
\end{figure}

\changed{Figure~\ref{fig:comparison_bars} shows the effect of different settings for the numbers of motion particles and sensors. We evaluated these changes by comparing the relative distances of three sets of animations (sitting, lying, walking) each containing five interactions. The descriptor is most effective when it differentiates interactions from the same group from those of different groups. Therefore, we measure the distance of interactions within the same group (internal) and across different groups (external). The best results were achieved using 2,000 motion particles and sensors generated by an Octree of level 7 or 8, which corresponds to about 3k active sensors for a domain size of 6 and an object sample distance of 0.1. Further increasing the number of the particles and sensors did not provide significantly better results. Additionally, we measured the effectiveness of using an Octree compared to a uniform grid with different cell sizes.}

\tog{\textbf{Comparison.} To evaluate our method, we used the \changed{\textit{Interaction Bisector Surface (IBS)}~\cite{Zhao:2014:ISU:2631978.2574860} and the more recently introduced \textit{Interaction Context (ICON)}~\cite{Hu:2015:ICT:2809654.2766914} descriptors that both describe interactions of groups of spatially arranged shapes by measuring the geometric properties of intersection surfaces between a center object and each of its surrounding shapes.}} ICON \changed{additionally provides \textit{Interaction Regions (IR)}} to facilitate functionality descriptors on these datasets that allow to classify shapes based on their function. However, \changed{both} approaches operate only on static scenes of two or more participating shapes, which makes a direct comparison with our method difficult.

\changed{To compare IBS and} ICON with our method we selected sets of animations and simulations in our dataset and converted them to a series of static scenes, by selecting snapshots of the motions. For each series, we selected pairs of motion driver and object, where the motion driver is closest to the surface of the observed shape and used this configuration as the input\erase{for ICON}. \changed{We tested this approach on six categories: bags,  bottles, and cups (grasping); chairs, benches, and beds (animations) which are composed of 65 time-variant scenes. Figure~\ref{fig:pr_diagrams} shows the precision and recall diagram. Our method allows to retrieve objects more reliably than other descriptors. Moreover, it returns objects that share the same functionality, even if their geometry differs.}

We also used the light-field descriptor (LFD)~\cite{CGF:CGF669} to evaluate our method as it provides a baseline for visual similarity based shape retrieval. It determines the similarity of two given shapes by their visual appearance. However, due to the semantic gap between geometry function, shapes with the same functionality potentially show large geometric variations. Consequently, the performance of LFD is inferior compared to our method, with a precision under 50\% for a recall of 60\% (Figure~\ref{fig:pr_diagrams}). Moreover, our method is capable of retrieving objects that have similar functionality, which LFD cannot reproduce.

\begin{figure}[t]
\centering
\includegraphics[width=0.98\linewidth]{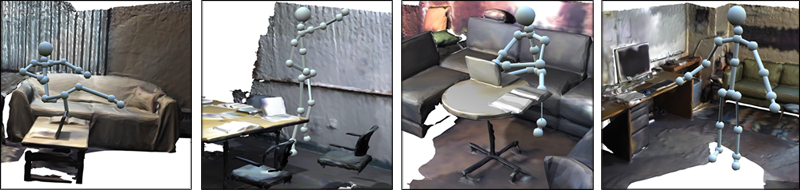}\hfill
\vspace{2mm}
\includegraphics[width=0.98\linewidth]{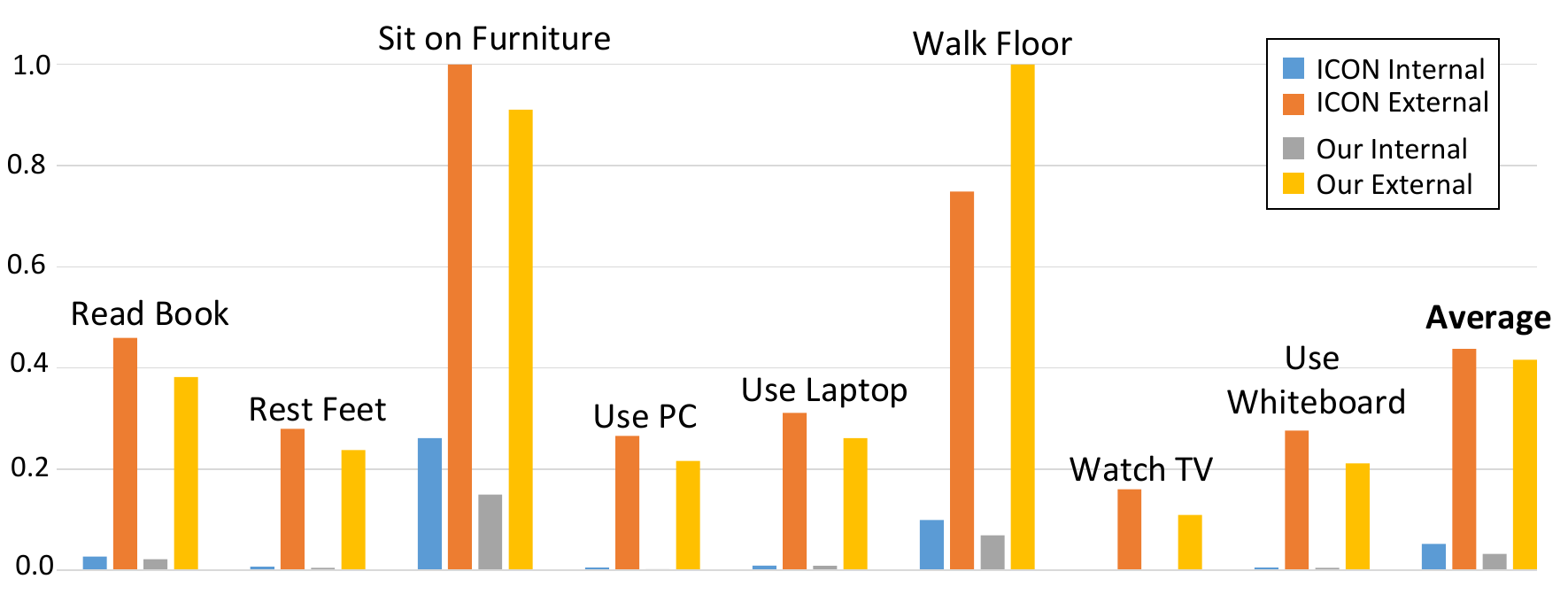}\hfill
\caption{A set of interactions from SceneGrok showing the interaction of a human with an environment (top). Compared to ICON our method produces similar results with respect to distances across different classes of interactions (external) while it performs better in discerning interactions of the same class (internal).}\label{fig:scenegrok_diagram}
\vspace{-4mm}
\end{figure}

\changed{
Figure~\ref{fig:scenegrok_diagram} shows the evaluation of our method and ICON on the RGB-D scans from Savva et al.~\shortcite{Savva:2014:SIA:2661229.2661230}. The dataset includes labeled interactions of a human in an environment captured from RGB-D scans. We use the provided motion information (captured skeleton) to build a stick figure mesh to sample and advect motion particles. The reconstructed environment mesh represents the observed object. We capture each type of interaction for the duration provided in the dataset. For ICON we used the figure mesh to compute the IBS and RS with the environment mesh.

Although our descriptor does not produce significantly different distances compared to ICON, it emphasizes the difference between interactions of the same groups (internal) and across different groups (external), with a ratio of 8.1 (ICON) to 11.2 (our) for the average distance. However, processing RGB-D scans poses a challenge in that they contain a lot of noise; the location of skeleton joints often jumps from frame to frame. \tog{\togerase{To avoid capturing these artifacts with our descriptor, we manually removed certain frames} To constrain these artifacts we only use sequences close to the interaction without considering transitions between them.} This explains the similarity to ICON as many of the interaction (\eg \textit{Read Book} or \textit{Rest Feet}) do not show significant amounts of motion. In case reliable motion information is available (\eg \textit{Walk Floor}) our descriptor provides better results.
}

\begin{figure}[t]
\includegraphics[width=\columnwidth]{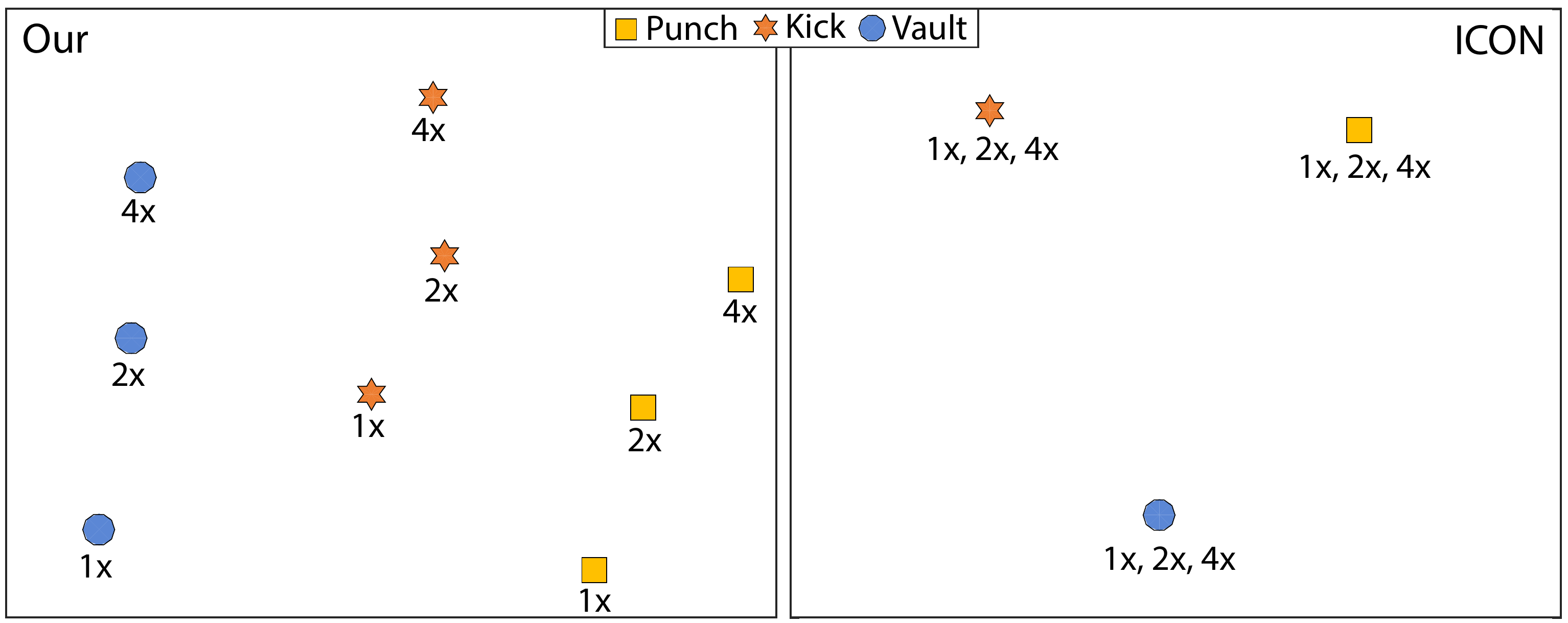}\hfill
\includegraphics[width=\columnwidth]{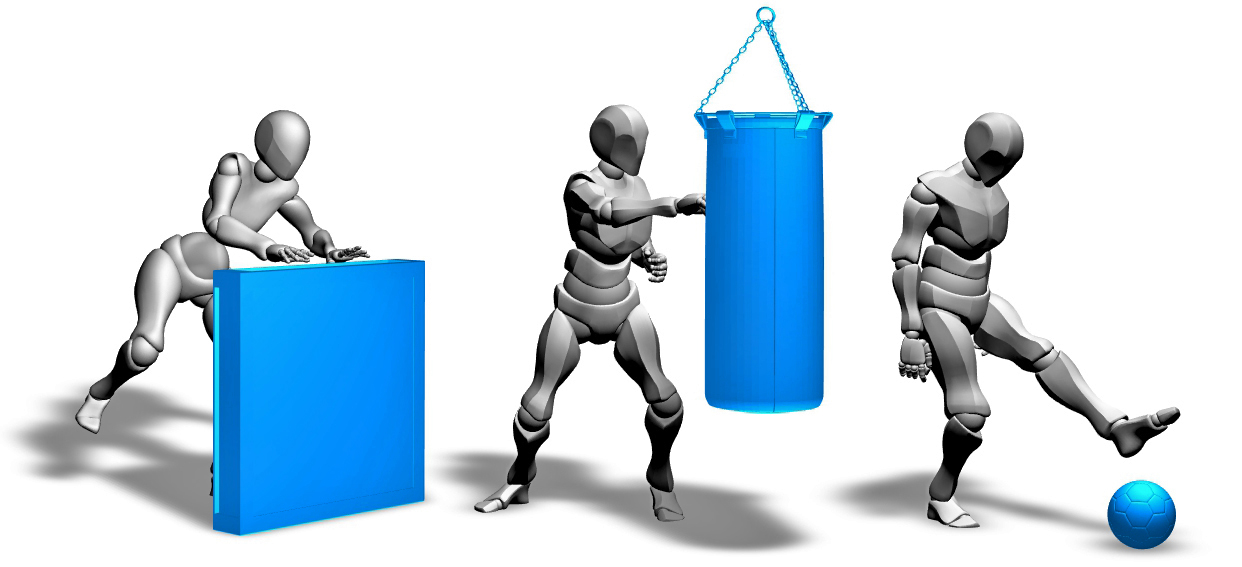}\hfill
\vspace{-2mm}
\caption{Top row: multidimensional scaling (MDS) plots of our method~(left) and ICON (right). While ICON can only differentiate the types of interactions, our method also allows to discriminate them according to their speed. Bottom row: we used three different interactions for this experiment at different speeds (1x, 2x, 4x).}
\label{fig:speed_comparison_icon}
\vspace{4mm}
\includegraphics[width=\columnwidth]{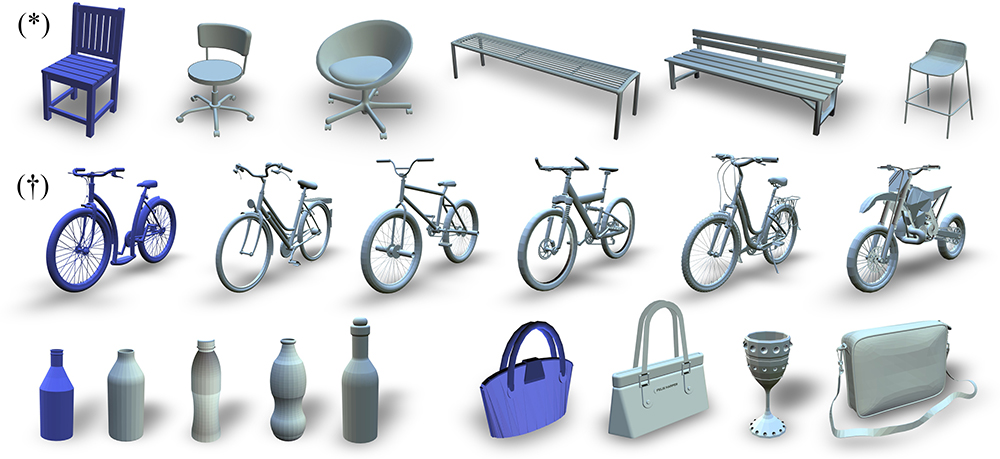}\hfill
\caption{Visual comparison of models retrieved by applying our descriptor. The query object is shown in blue, the closest models according to our descriptor to the right; the retrieved objects share similar properties in descending order. The motion drivers used for this experiment were: sitting animation (chairs), fluid simulation (bicycles), grasping simulation (bottles, bags).}
\label{fig:shape_retrieval}
\vspace{-2mm}
\end{figure}

\tog{Figure~\ref{fig:speed_comparison_icon} illustrates the advantage of exploiting motions for discriminating interactions. We captured three animations of humanoid meshes interacting with static objects at different speeds~(bottom row). As can be seen in the embedding space~(top, left), our method allows to discern the type of interaction~(punch, kick, vault) as well as variations in the speed at which they are performed. Other descriptors do not consider speed as feature, which is illustrated in Figure~\ref{fig:speed_comparison_icon} (top, right). We used ICON for an evaluation on our datasets by taking static snapshots of the motion driver sampled at relative temporal distances of the animation. Each static snapshot is a similar object-motion arrangement, independent of the variations in speed. Hence, ICON can differentiate the type of interaction, but not the variations of the motion. The MDS plot on the right visualizes the same interactions shown on the left. All static snapshots of the same interaction appear similar to ICON and are thus positioned at the same location in the embedding space.} 

Finally, it is important to note that our descriptor, although designed for motion comparison, provides comparable results to descriptors that are designed for static scenes.

\subsection{Applications}
\label{sec:applications}

To demonstrate the effectiveness of our interaction descriptor we illustrate its application for shape classification and saliency estimation. \tog{In particular, we perform tests on shape retrieval by querying objects from a dataset, by estimating salient regions of shapes based on their interaction response, and by analyzing motions to predict the corresponding type of interaction.} \changed{It is important to note that we expect shapes of the same class to share a common orientation and require a canonical arrangement of the interaction landscape with respect to the observed shape. Objects can store signatures of multiple interactions with various motion drivers.}

\textbf{Shape Retrieval.} We quantify the performance based on the quality of shape retrievals from a dataset containing groups of labeled shapes \changed{(157 objects, 20 classes). Each object in the database stores a signature of the interaction with a motion driver that we manually assigned to each class. We select an unknown query object and perform an interaction with the same motion driver. Finally, we determine the precision and recall of shapes in the database by searching for the closest matches using the relative distance (Section~\ref{sec:encoding_motion_flows}) and by counting true and false positives until we recall all shapes from a given class.} Figure~\ref{fig:shape_retrieval} shows the result for some of the objects used for this experiment. We query our dataset with an input shape (blue), \changed{right to it we see the returned shapes in prioritized order (from left to right). Except for the failure case of the wineglass, our method returns semantically similar shapes (an object is a chair because you use it as a chair). \tog{Unlike the current approaches for describing static object relationships, capturing interactions as continuous motions implicitly reveals the affordance of the observed shapes in a more principled way.} Although we rely on the specific arrangement of object and motion driver, this experiment emphasizes the potential of our method, as we are able to retrieve shapes that share similarities in their form and function.} 


\textbf{Saliency Estimation.} Shapes are often synthesized without context and without knowledge about the semantic identity of their parts, making predictions on their functional use a difficult problem in structure-aware modeling.~Previous methods proposed finding salient features of shapes through analyzing their geometric representation~\cite{Gal:2006:SGF:1122501.1122507,Shilane:2007:DRS:1243980.1243981}, however, they only operate on shapes in isolation. Kim et al.~\shortcite{Kim:2014:SHS:2601097.2601117} proposed to predict salient features by finding contact points of human agents interacting with shapes; their method explicitly focuses on human-centric salience estimation.

\begin{figure}[t]
\includegraphics[width=\columnwidth]{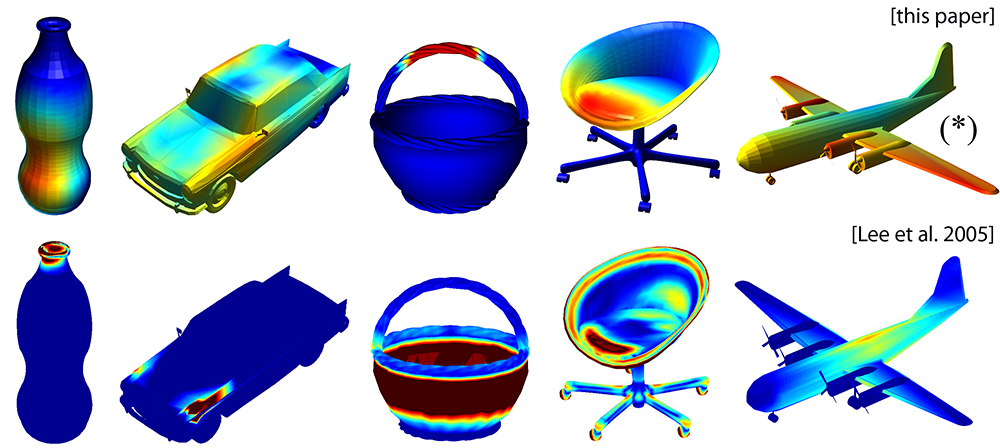}\hfill
\caption{A comparison of surface salience estimations of our approach \erase{(bottom)}\changed{(top)} and Lee et al.~[2005] \erase{(top)}\changed{(bottom)}. Salient regions are highlighted as color overlay, red indicates a high response to the performed interaction, blue identifies uninvolved areas. \changed{The motion drivers for this experiment were: side grasp (flask), wind (car), top grasp (basket), animation (chair), wind (plane).}}
\label{fig:saliency_estimation}
\vspace{6mm}
\includegraphics[width=\columnwidth]{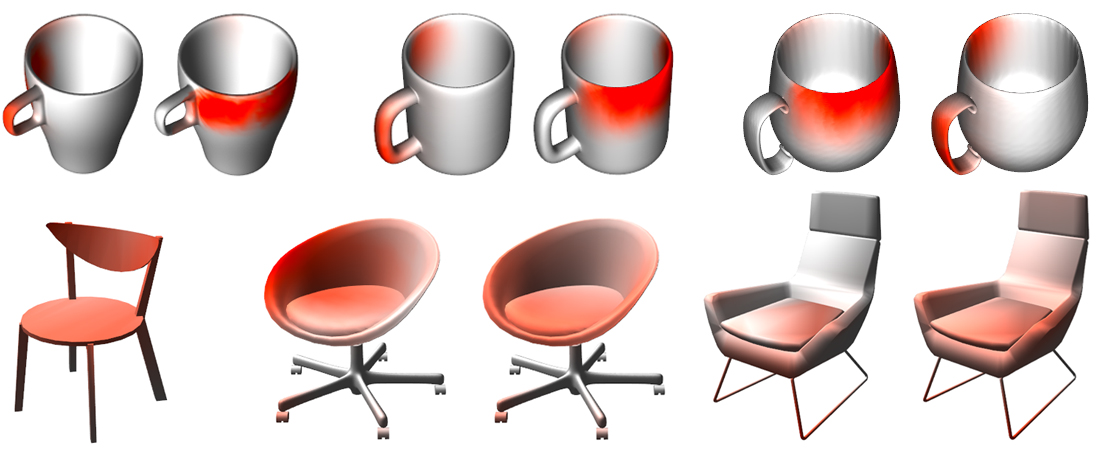}\hfill
\hspace*{6mm}(a)\hspace*{16mm}(b)\hspace*{14mm}(c)\hspace*{12mm}(d)\hspace*{13mm}(e)\hfill~
\vspace{1mm}
\caption{An example of matching correspondences with our descriptor. We can apply the same interaction to different shapes and find their their corresponding regions if saliency is shared among the shapes. The top row shows three cups with color coded corresponding regions, handle (grasp from the side) and rim (grasp from top). The bottom row shows the result of an interaction transfer experiment. We capture a sitting interaction (animation) for the three chairs (a), (b), and (d). The captured interaction of (a) is transfered to the other chairs (c), (e).}
\label{fig:matching_correspondences}
\end{figure}

We track interactions as continuous events in the interaction space of two shapes. We not only know where the motion driver is in direct contact with the observed shape, but also how it approaches its surface, or even how it initiates the interaction. \erase{We exploit this knowledge to determine salient features of shapes based on their functionality.}We estimate salient regions by mapping the motion signature data to the geometry of a model. We find the sensor region for each \changed{shape} vertex and its corresponding vector field. \changed{The saliency is measured for each vertex of the observed shape by computing the weighted sum of vector field attributes of the cell that contains the vertex. The weight for each attribute was found through experiments (Section \ref{sec:encoding_motion_flows}). Additionally, we define a spherical region around the center of the current cell and include all adjacent cells with a distance smaller than the specified radius in the computation. We assign the normalized weighted sum to each vertex as the saliency measure.} Figure~\ref{fig:saliency_estimation} (top) shows a visualization of salient regions in direct comparison with the surface curvature based approach of Lee et al.~\shortcite{Lee:2005:MS:1073204.1073244} (bottom). Although we do not conclude that our saliency estimation provides better results, we would like to emphasize that an interaction-based estimation is a more general means that provides meaningful hints towards high level saliency detection. \tog{Compared to existing methods that solely rely on analyzing the geometry for finding salient regions, exploiting interaction information provides more details of the observed object to the cost of a more complex setup}.  

\textbf{Shape Correspondence.} Matching correspondences of semantically similar parts remains a challenging problem in many areas, such as shape completion, and segmentation, similarity assessment, or structure-aware modeling. \changed{To test our descriptor on this task, we find salient regions of shapes of the same category and determine the sum of their differences. Vertices of participating meshes are inserted into a uniform grid and the saliency value assigned to each vertex is used for the computation.} Regions with the minimal sum of these differences can be defined as corresponding regions. This is similar to what Kim et al.~\shortcite{Kim:2014:SHS:2601097.2601117} define for human-centric interactions. We can identify multiple interactions on the same object (\eg grasping a cup from above or from the side). By finding the best match between these two kind of interaction signatures, our method allows us to detect corresponding regions of geometrically different shapes. Figure~\ref{fig:matching_correspondences} illustrates the results between a set of cups. \changed{The bottom row shows the result of an interaction transfer experiment. Here we capture a sitting interaction (animation) for the three chairs (a), (b), and (d). The results of (a) is transfered to the other chairs (c), (e).}

\begin{figure}[t]
\includegraphics[width=0.5\columnwidth]{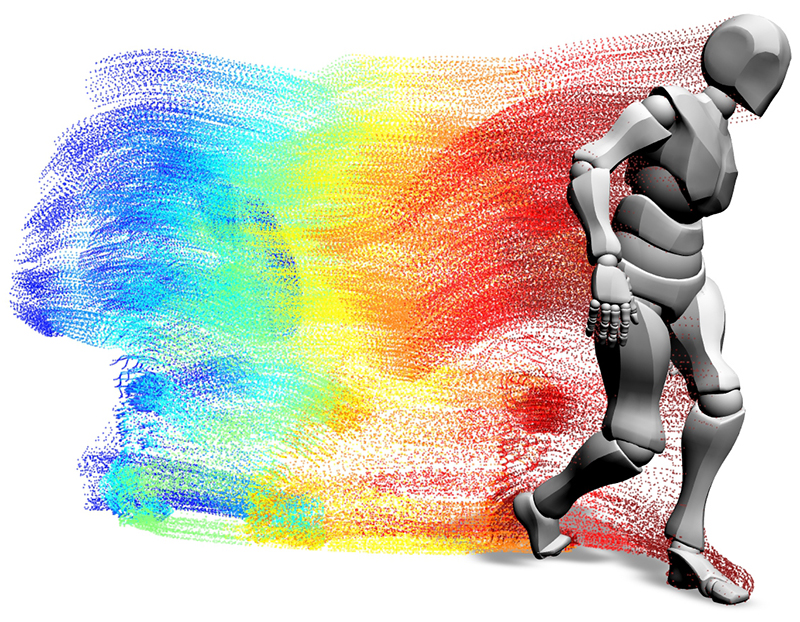}\hfill
\includegraphics[width=0.5\columnwidth]{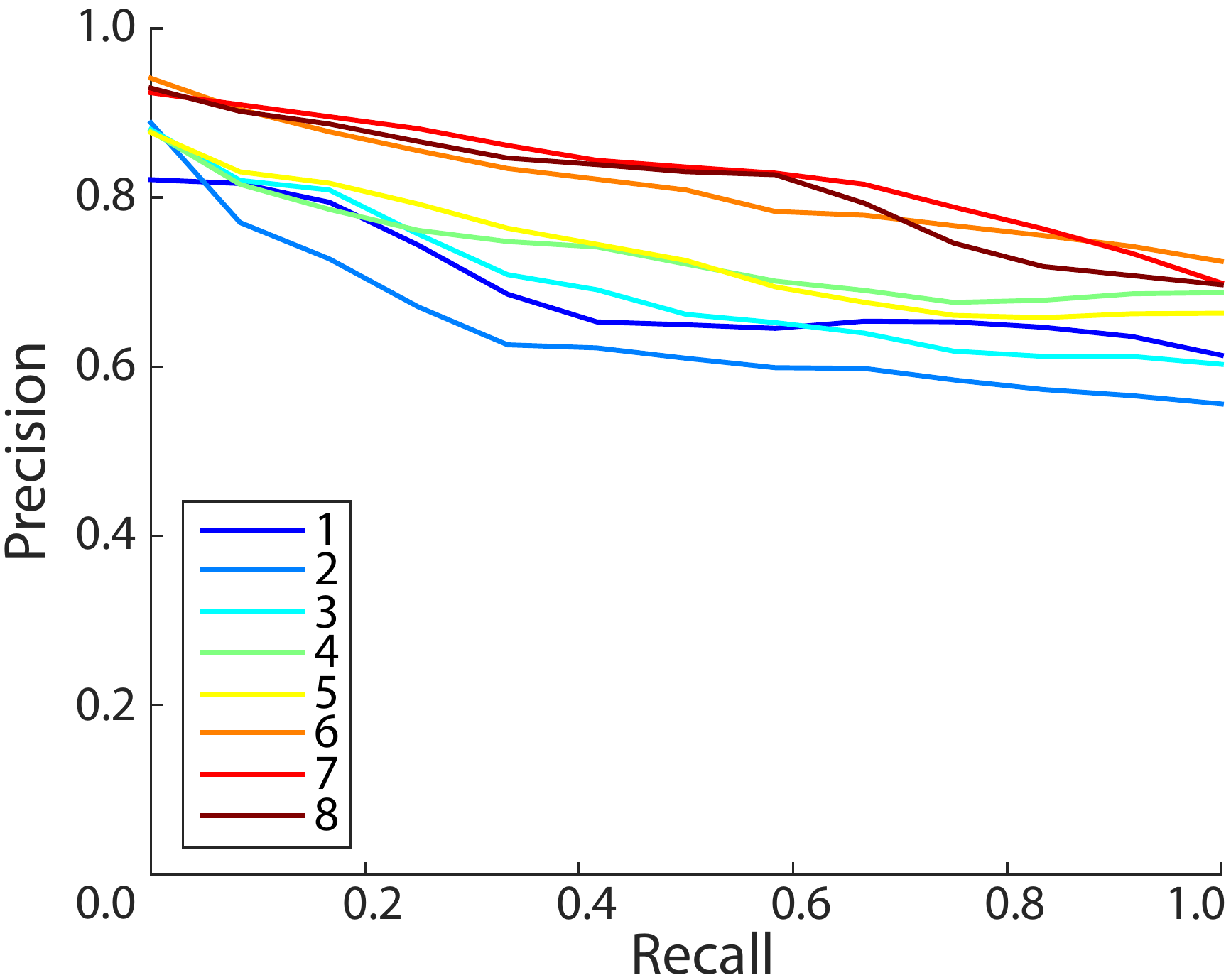}\hfill
\caption{Interaction prediction: we divide interactions into equidistant time spans (left, start: blue, end: red) and store the signature for each part in a database. We predict an interaction by  progressively comparing sequences of an unseen interaction to the previously encoded segments in the database. Right: a PR diagram that visualizes the quality of the interaction prediction. More sequences allow to more reliably predict an interaction. The color of the curves in the PR diagram matches the color of the sequence up to where we compare the interaction.}
\label{fig:motion_prediction}
\vspace{-4mm}
\end{figure}

\textbf{Interaction Prediction.}
\tog{
Finally, we show that exploiting time variations of interactions enables new types of applications that cannot be supported by static interaction descriptors. Our method can be used to predict the type of an interaction when only the first few seconds are known. 
To perform this experiment, we captured interactions and divided the motion trajectories into $N=8$ time spans, for which we independently compute a signature. The k'th signature of an interaction is computed based on the trajectories captured between the beginning $t_{begin}$ and $t_{begin}+\frac{k}{N}d$, where $d$ is the duration of the interaction. The set of signatures for each interaction is stored in a database. We used 23 interactions of 3 categories to produce a total of 184~signatures. To perform the prediction of an unseen interaction we progressively calculate the signature of the already available part of an interaction and compare it to the signatures in the database that cover the same number of time spans.

The performance of predicting the type of an interaction is visualized in Figure~\ref{fig:motion_prediction}. On the left we show a walking motion with color-coded trajectories. Each color represents a temporal sequence of the interaction. The PR diagram on the right shows how the performance gradually increases the more time spans become available. The graphs show the average precision and recall values for all interactions used in this experiment. With only one segment available we achieve a precision of 60\% with a recall of 50\%. The color in the PR diagram matches the color of the corresponding sequence up to where we compute the signature for the interaction. Current descriptors for interactions provide a means for analyzing and predicting the functionality of objects, but do not allow to predict interactions as dynamic motions. Unlike the previous approaches, our descriptor analyzes the motion flow and allows to reliably predict the type of interaction. 
}

\vspace{-2mm}
\subsection{Discussion and Limitations}

The main limitation of our application is the absence of more meaningful interaction data. Predefined animations do not provide the same complexity in the motions found in the real-world or simulated data sets. Moreover, some of our current observations rely on small groups of predefined data resulting in a sparse representation of the captured interactions. It would be interesting to overcome this limitation and to test our method with more data from RGB-D video sequences of real-world interactions.

Unlike \changed{IBS~\cite{Zhao:2014:ISU:2631978.2574860}} and ICON~\cite{Hu:2015:ICT:2809654.2766914}, we do not provide a hierarchical aggregation of interactions, but only consider the entire interaction landscape of the observed object, which could yield less meaningful signatures for complex interaction of many objects. We explicitly limit ourselves to a single object interacting with one or more motion drivers and would like to address more complex interaction concepts as future work. Moreover, our method is constrained by the need to place motion particles on moving surfaces. While this does not pose a problem for Lagrangian simulations, we rely on an expensive sampling of surfaces to find motion particle positions for solid objects. Moreover, only sampling a surface causes imprecision in capturing the motion trajectories and the corresponding interaction. However, capturing interactions through tracking complex geometric surfaces or even volumes does not yet seem feasible.

\tog{Instead of employing vector fields to quantize motions, potentially more refined representation exist. For example,  directly defining and exploiting properties on the motion trajectories could be an interesting endeavor. However, to our knowledge, no method exists to readily explore such attributes of time-variant trajectories as geometric entities. Just finding correspondences or common parts within a single trajectory is a difficult problem that cannot be solved easily and at the required scale. Compared to more lightweight approaches that directly operate on sensor data, employing vector fields is less prone to artifacts and noise in the input data. The vector fields regularize the signal of the motion flow and can be efficiently analyzed. Our descriptor is based on six attributes and allows us to encode finer nuances of the motions compared to, \eg capturing the properties of normals of propagating surfaces as discussed in Oreifej and Liu~\shortcite{6618942}. 
}
\section{Conclusion and Future Work}
We have presented a novel general-purpose descriptor that exploits time-variant properties of interactions observed in 3-dimensional spaces. We define interaction landscapes as the interaction space of two objects participating in an interaction. We distribute sensors in this interaction space and place motion particles on moving surfaces to track their movements. Our framework allows capturing interactions from a variety of sources, including simulations, animations, \changed{motion capture, and RGB-D scans}. By encoding the resulting motion trajectories of the tracked particles, we can acquire information that supports a more complete understanding of shapes and their function. Moreover, we have shown that the signatures of observed interactions can be used to classify shapes and to find their salient regions.

Capturing the interaction of objects is essential in order to learn more about shapes and their function. Moreover, we believe that a general purpose descriptor provides a valuable means for exploring and classifying interactions with many potential avenues of future work. In this work we specifically limited ourselves to encoding static interaction landscapes of one motion driver and an observed object. However, in future work observing objects in dynamically changing interaction spaces seems like an interesting endeavor. Furthermore, the interactions of entire groups of entities, \eg as can be found in crowds or swarms, \hide{also on larger scales,} could be explored.

\tog{An interesting avenue for future work would be to explore our descriptor for other, non-proximal, interaction types as many interactions do not happen through object-object contacts, but remotely.} This spans from physiological phenomena, such as capturing views or non-verbal human behavior to non-contact physical interactions, such as the transport of light or the effect of magnetism. Our method provides a general means for encoding interactions applicable to many domains. Finally, we would like to explore more elaborate methods to better encode finer nuances in the captured motion flows and to thereby enable our descriptor for more evolved applications in shape analysis, such as functional correspondence, shape synthesis, and symmetry detection.

\vspace{-2mm}
\section*{Acknowledgements}
\tog{We would like to thank Mirela Ben-Chen for her insightful comments and suggestions on methods for 3D vector fields and Torsten H\"adrich for his implementation of the SPH-based fluid solver. The work was supported by the NSF grants CCF-1514305, IIS-1528025, EEC-1606396, NSG-1161480, the Stanford AI Lab-Toyota Center for Artificial Intelligence Research, a Google Focused Research Award, the Max Planck Center for Visual Computing and Communications, the JSPS Strategic Young Researchers Visits Program for Acceleration Brain Circulations, and the National Science Foundation of China (61373071). 
}

\bibliographystyle{acmsiggraph}
\bibliography{main}

\begin{thebibliography}{\protect\citename{Karlsruhe Institue~of Technology
  }2015}

\bibitem[\protect\citename{Adobe }2015]{mixamo:2015}
{\sc Adobe}, 2015.
\newblock Mixamo.
\newblock \url{https://www.mixamo.com/}.

\bibitem[\protect\citename{Al-Asqhar et~al\mbox{.}
  }2013]{Al-Asqhar:2013:RDI:2485895.2485905}
{\sc Al-Asqhar, R.~A., Komura, T., and Choi, M.~G.}
\newblock 2013.
\newblock Relationship descriptors for interactive motion adaptation.
\newblock In {\em Proc. of ACM SIGGRAPH/Eurographics Symposium on Computer
  Animation}, ACM, SCA '13, 45--53.

\bibitem[\protect\citename{Amenta et~al\mbox{.}
  }1998]{Amenta:1998:NVS:280814.280947}
{\sc Amenta, N., Bern, M., and Kamvysselis, M.}
\newblock 1998.
\newblock A new voronoi-based surface reconstruction algorithm.
\newblock In {\em Proc. of SIGGRAPH}, ACM, New York, NY, USA, 415--421.

\bibitem[\protect\citename{Anguelov et~al\mbox{.} }2005]{Anguelov:SIGGRAPH05}
{\sc Anguelov, D., Srinivasan, P., Koller, D., Thrun, S., Rodgers, J., and
  Davis, J.}
\newblock 2005.
\newblock {SCAPE}: {S}hape completion and animation of people.
\newblock In {\em Proc. of SIGGRAPH}.

\bibitem[\protect\citename{Ashby }1992]{ashby1992multidimensional}
{\sc Ashby, F.}
\newblock 1992.
\newblock {\em Multidimensional Models of Perception and Cognition}.
\newblock Lea's Communication. L. Erlbaum.

\bibitem[\protect\citename{Bar-Aviv and Rivlin }2006]{Bar-AvivR06}
{\sc Bar-Aviv, E., and Rivlin, E.}
\newblock 2006.
\newblock Functional 3d object classification using simulation of embodied
  agent.
\newblock In {\em BMVC}, British Machine Vision Association, 307--316.

\bibitem[\protect\citename{Baran and Popovi{\'{c}} }2007]{Baran:2007:ARA}
{\sc Baran, I., and Popovi{\'{c}}, J.}
\newblock 2007.
\newblock Automatic rigging and animation of 3{D} characters.
\newblock {\em ACM Trans. Graph. 26}, 3 (jul), 72:1--72:8.

\bibitem[\protect\citename{Biasotti et~al\mbox{.} }2015]{CGF:CGF12734}
{\sc Biasotti, S., Cerri, A., Bronstein, A., and Bronstein, M.}
\newblock 2015.
\newblock Recent trends, applications, and perspectives in 3d shape similarity
  assessment.
\newblock {\em Comp. Graph. Forum\/}, n/a--n/a.

\bibitem[\protect\citename{Black and Taylor
  }1997]{Black97automaticallyclustering}
{\sc Black, A.~W., and Taylor, P.}
\newblock 1997.
\newblock Automatically clustering similar units for unit selection in speech
  synthesis.
\newblock In {\em in Eurospeech97}, 601--604.

\bibitem[\protect\citename{Caine }1994]{351268}
{\sc Caine, M.}
\newblock 1994.
\newblock The design of shape interactions using motion constraints.
\newblock In {\em Robotics and Automation, 1994. Proceedings., 1994 IEEE
  International Conference on}, 366--371 vol.1.

\bibitem[\protect\citename{{Chang} et~al\mbox{.} }2015]{2015arXiv151203012C}
{\sc {Chang}, A.~X., {Funkhouser}, T., {Guibas}, L., {Hanrahan}, P., {Huang},
  Q., {Li}, Z., {Savarese}, S., {Savva}, M., {Song}, S., {Su}, H., {Xiao}, J.,
  {Yi}, L., and {Yu}, F.}
\newblock 2015.
\newblock {ShapeNet: An Information-Rich 3D Model Repository}.
\newblock {\em ArXiv e-prints\/} (Dec.).

\bibitem[\protect\citename{Chao et~al\mbox{.} }2015]{Chao2015}
{\sc Chao, Y.-W., Wang, Z., He, Y., Wang, J., and Deng, J.}
\newblock 2015.
\newblock Hico: A benchmark for recognizing human-object interactions in
  images.
\newblock {\em ICCV\/}.

\bibitem[\protect\citename{Chen et~al\mbox{.} }2003]{CGF:CGF669}
{\sc Chen, D.-Y., Tian, X.-P., Shen, Y.-T., and Ouhyoung, M.}
\newblock 2003.
\newblock On visual similarity based 3d model retrieval.
\newblock {\em Comp. Graph. Forum 22}, 3, 223--232.

\bibitem[\protect\citename{Chen et~al\mbox{.}
  }2013]{Chen:2013:BBN:2508363.2508375}
{\sc Chen, J., Ge, X., Wei, L.-Y., Wang, B., Wang, Y., Wang, H., Fei, Y., Qian,
  K.-L., Yong, J.-H., and Wang, W.}
\newblock 2013.
\newblock Bilateral blue noise sampling.
\newblock {\em ACM Trans. Graph. 32}, 6 (Nov.), 216:1--216:11.

\bibitem[\protect\citename{Durrleman }2010]{Durleman2010}
{\sc Durrleman, S.}
\newblock 2010.
\newblock {\em Statistical models of currents for measuring the variability of
  anatomical curves, surfaces and their evolution}.
\newblock PhD thesis, Universit\'{e} Nice - Sophia Antipolis, France.

\bibitem[\protect\citename{Fisher et~al\mbox{.}
  }2015]{Fisher:2015:ASS:2816795.2818057}
{\sc Fisher, M., Savva, M., Li, Y., Hanrahan, P., and Niessner, M.}
\newblock 2015.
\newblock Activity-centric scene synthesis for functional 3d scene modeling.
\newblock {\em ACM Trans. Graph. 34}, 6 (Oct.), 179:1--179:13.

\bibitem[\protect\citename{Gal and Cohen-Or
  }2006]{Gal:2006:SGF:1122501.1122507}
{\sc Gal, R., and Cohen-Or, D.}
\newblock 2006.
\newblock Salient geometric features for partial shape matching and similarity.
\newblock {\em ACM Trans. Graph. 25}, 1, 130--150.

\bibitem[\protect\citename{Grabner et~al\mbox{.} }2011]{5995327}
{\sc Grabner, H., Gall, J., and Van~Gool, L.}
\newblock 2011.
\newblock What makes a chair a chair?
\newblock In {\em CVPR}, 1529--1536.

\bibitem[\protect\citename{Gupta et~al\mbox{.}
  }2009]{Gupta:2009:OHI:1608576.1608766}
{\sc Gupta, A., Kembhavi, A., and Davis, L.~S.}
\newblock 2009.
\newblock Observing human-object interactions: Using spatial and functional
  compatibility for recognition.
\newblock {\em IEEE Trans. Pattern Anal. Mach. Intell. 31}, 10 (Oct.),
  1775--1789.

\bibitem[\protect\citename{Ho et~al\mbox{.} }2010]{Ho:2010:SRP:1778765.1778770}
{\sc Ho, E. S.~L., Komura, T., and Tai, C.-L.}
\newblock 2010.
\newblock Spatial relationship preserving character motion adaptation.
\newblock {\em ACM Trans. Graph. 29}, 4, 33:1--33:8.

\bibitem[\protect\citename{Hu et~al\mbox{.} }2015]{Hu:2015:ICT:2809654.2766914}
{\sc Hu, R., Zhu, C., van Kaick, O., Liu, L., Shamir, A., and Zhang, H.}
\newblock 2015.
\newblock Interaction context (icon): Towards a geometric functionality
  descriptor.
\newblock {\em ACM Trans. Graph. 34}, 4, 83:1--83:12.

\bibitem[\protect\citename{Hu et~al\mbox{.} }2016]{Hu:2016:LOF:2897824.2925870}
{\sc Hu, R., van Kaick, O., Wu, B., Huang, H., Shamir, A., and Zhang, H.}
\newblock 2016.
\newblock Learning how objects function via co-analysis of interactions.
\newblock {\em ACM Trans. Graph. 35}, 4 (July), 47:1--47:13.

\bibitem[\protect\citename{Jiang and Martin }2008]{jiang2008}
{\sc Jiang, H., and Martin, D.~R.}
\newblock 2008.
\newblock Finding actions using shape flows.
\newblock In {\em Computer Vision – ECCV 2008}, D.~Forsyth, P.~Torr, and
  A.~Zisserman, Eds., vol.~5303 of {\em Lecture Notes in Computer Science}.
  Springer Berlin Heidelberg, 278--292.

\bibitem[\protect\citename{Karlsruhe Institue~of Technology
  }2015]{HumanMotionDB:2015}
{\sc Karlsruhe Institue~of Technology, Karlsruhe, G.}, 2015.
\newblock Kit whole human motion database.
\newblock \url{https://motion-database.humanoids.kit.edu/}.

\bibitem[\protect\citename{Kim et~al\mbox{.}
  }2014]{Kim:2014:SHS:2601097.2601117}
{\sc Kim, V.~G., Chaudhuri, S., Guibas, L., and Funkhouser, T.}
\newblock 2014.
\newblock Shape2pose: human-centric shape analysis.
\newblock {\em ACM Trans. Graph. 33}, 4, 120:1--120:12.

\bibitem[\protect\citename{Laga et~al\mbox{.}
  }2013]{Laga:2013:GCS:2516971.2516975}
{\sc Laga, H., Mortara, M., and Spagnuolo, M.}
\newblock 2013.
\newblock Geometry and context for semantic correspondences and functionality
  recognition in man-made 3d shapes.
\newblock {\em ACM Trans. Graph. 32}, 5, 150:1--150:16.

\bibitem[\protect\citename{Lee et~al\mbox{.}
  }2005]{Lee:2005:MS:1073204.1073244}
{\sc Lee, C.~H., Varshney, A., and Jacobs, D.~W.}
\newblock 2005.
\newblock Mesh saliency.
\newblock {\em ACM Trans. Graph. 24}, 3 (July), 659--666.

\bibitem[\protect\citename{Li et~al\mbox{.} }2007]{4293017}
{\sc Li, Y., Fu, J.~L., and Pollard, N.~S.}
\newblock 2007.
\newblock Data-driven grasp synthesis using shape matching and task-based
  pruning.
\newblock {\em IEEE Transactions on Visualization and Computer Graphics 13}, 4
  (July), 732--747.

\bibitem[\protect\citename{Li et~al\mbox{.} }2015]{Li:2015:QCM:2870647.2753755}
{\sc Li, P., Wang, B., Sun, F., Guo, X., Zhang, C., and Wang, W.}
\newblock 2015.
\newblock Q-mat: Computing medial axis transform by quadratic error
  minimization.
\newblock {\em ACM Trans. Graph. 35}, 1 (Dec.), 8:1--8:16.

\bibitem[\protect\citename{Liu et~al\mbox{.} }2015]{Liu2015110}
{\sc Liu, Z., Xie, C., Bu, S., Wang, X., Han, J., Lin, H., and Zhang, H.}
\newblock 2015.
\newblock Indirect shape analysis for 3d shape retrieval.
\newblock {\em Computers \& Graphics 46\/}, 110--116.
\newblock Shape Modeling International 2014.

\bibitem[\protect\citename{Mitra et~al\mbox{.}
  }2013a]{Mitra:2013:SSP:2542266.2542267}
{\sc Mitra, N., Wand, M., Zhang, H.~R., Cohen-Or, D., Kim, V., and Huang,
  Q.-X.}
\newblock 2013.
\newblock Structure-aware shape processing.
\newblock In {\em SIGGRAPH Asia 2013 Courses}, ACM, New York, NY, USA, SA '13,
  1:1--1:20.

\bibitem[\protect\citename{Mitra et~al\mbox{.} }2013b]{journals/cgf/MitraPWC13}
{\sc Mitra, N.~J., Pauly, M., Wand, M., and Ceylan, D.}
\newblock 2013.
\newblock Symmetry in 3d geometry: Extraction and applications.
\newblock {\em Comput. Graph. Forum 32}, 6, 1--23.

\bibitem[\protect\citename{Monaghan }1992]{monaghan1992smoothed}
{\sc Monaghan, J.~J.}
\newblock 1992.
\newblock Smoothed particle hydrodynamics.
\newblock {\em Annual review of astronomy and astrophysics 30\/}, 543--574.

\bibitem[\protect\citename{Oreifej and Liu }2013]{6618942}
{\sc Oreifej, O., and Liu, Z.}
\newblock 2013.
\newblock Hon4d: Histogram of oriented 4d normals for activity recognition from
  depth sequences.
\newblock In {\em Computer Vision and Pattern Recognition (CVPR), 2013 IEEE
  Conference on}, 716--723.

\bibitem[\protect\citename{Osada et~al\mbox{.}
  }2002]{Osada:2002:SD:571647.571648}
{\sc Osada, R., Funkhouser, T., Chazelle, B., and Dobkin, D.}
\newblock 2002.
\newblock Shape distributions.
\newblock {\em ACM Trans. Graph. 21}, 4 (Oct.), 807--832.

\bibitem[\protect\citename{Ovsjanikov et~al\mbox{.} }2012]{fmapssig2012}
{\sc Ovsjanikov, M., Ben-Chen, M., Solomon, J., Butscher, A., and Guibas, L.}
\newblock 2012.
\newblock Functional maps: A flexible representation of maps between shapes.
\newblock {\em ACM Trans. Graph. 31}, 4.

\bibitem[\protect\citename{Pechuk et~al\mbox{.}
  }2008]{Pechuk:2008:LFO:1363359.1363379}
{\sc Pechuk, M., Soldea, O., and Rivlin, E.}
\newblock 2008.
\newblock Learning function-based object classification from 3d imagery.
\newblock {\em Comput. Vis. Image Underst. 110}, 2 (May), 173--191.

\bibitem[\protect\citename{Rivlin et~al\mbox{.} }1994]{323839}
{\sc Rivlin, E., Dickinson, S., and Rosenfeld, A.}
\newblock 1994.
\newblock Recognition by functional parts [function-based object recognition].
\newblock In {\em Computer Vision and Pattern Recognition, 1994. Proceedings
  CVPR '94., 1994 IEEE Computer Society Conference on}, 267--274.

\bibitem[\protect\citename{Robotics~Lab }2012]{graspit:2012}
{\sc Robotics~Lab, Computer Science~Laboratory, C. U.~U.}, 2012.
\newblock Graspit!
\newblock \url{http://www.cs.columbia.edu/~cmatei/graspit/}.

\bibitem[\protect\citename{Rubner et~al\mbox{.} }1998]{710701}
{\sc Rubner, Y., Tomasi, C., and Guibas, L.}
\newblock 1998.
\newblock A metric for distributions with applications to image databases.
\newblock In {\em Computer Vision, 1998. Sixth International Conference on},
  59--66.

\bibitem[\protect\citename{Savva et~al\mbox{.}
  }2014]{Savva:2014:SIA:2661229.2661230}
{\sc Savva, M., Chang, A.~X., Hanrahan, P., Fisher, M., and Nie{\ss}ner, M.}
\newblock 2014.
\newblock Scenegrok: Inferring action maps in 3d environments.
\newblock {\em ACM Trans. Graph. 33}, 6 (Nov.), 212:1--212:10.

\bibitem[\protect\citename{Savva et~al\mbox{.} }2016]{Savva:2016}
{\sc Savva, M., Chang, A.~X., Hanrahan, P., Fisher, M., and Nie{\ss}ner, M.}
\newblock 2016.
\newblock Pigraphs: Learning interaction snapshots from observations.
\newblock {\em ACM Trans. Graph. 35}, 4 (July), 139:1--139:12.

\bibitem[\protect\citename{Shilane and Funkhouser
  }2007]{Shilane:2007:DRS:1243980.1243981}
{\sc Shilane, P., and Funkhouser, T.}
\newblock 2007.
\newblock Distinctive regions of 3d surfaces.
\newblock {\em ACM Trans. Graph. 26}, 2 (June).

\bibitem[\protect\citename{Sidi et~al\mbox{.}
  }2011]{Sidi:2011:UCS:2070781.2024160}
{\sc Sidi, O., van Kaick, O., Kleiman, Y., Zhang, H., and Cohen-Or, D.}
\newblock 2011.
\newblock Unsupervised co-segmentation of a set of shapes via descriptor-space
  spectral clustering.
\newblock {\em ACM Trans. Graph. 30}, 6, 126:1--126:10.

\bibitem[\protect\citename{Song et~al\mbox{.} }2011]{6130360}
{\sc Song, H.~O., Fritz, M., Gu, C., and Darrell, T.}
\newblock 2011.
\newblock Visual grasp affordances from appearance-based cues.
\newblock In {\em Computer Vision Workshops (ICCV Workshops), 2011 IEEE
  International Conference on}, 998--1005.

\bibitem[\protect\citename{Sutton et~al\mbox{.} }1994]{Sutton19941743}
{\sc Sutton, M., Stark, L., and Bowyer, K.}
\newblock 1994.
\newblock Gruff-3: Generalizing the domain of a function-based recognition
  system.
\newblock {\em Pattern Recognition 27}, 12, 1743 -- 1766.

\bibitem[\protect\citename{Tevs et~al\mbox{.}
  }2014]{Tevs:2014:RSV:2601097.2601220}
{\sc Tevs, A., Huang, Q., Wand, M., Seidel, H.-P., and Guibas, L.}
\newblock 2014.
\newblock Relating shapes via geometric symmetries and regularities.
\newblock {\em ACM Trans. Graph. 33}, 4 (July), 119:1--119:12.

\bibitem[\protect\citename{Tversky }1977]{FeaturesOfSimilarity.pdf}
{\sc Tversky, A.}
\newblock 1977.
\newblock Features of similarity.
\newblock {\em Psychological Review 84\/}, 327--352.

\bibitem[\protect\citename{Tzionas and Gall }2015]{Tzionas2015}
{\sc Tzionas, D., and Gall, J.}
\newblock 2015.
\newblock 3d object reconstruction from hand-object interactions.
\newblock In {\em International Conference on Computer Vision (ICCV)}.

\bibitem[\protect\citename{Wu et~al\mbox{.} }2014]{DBLP:journals/corr/WuSKTX14}
{\sc Wu, Z., Song, S., Khosla, A., Tang, X., and Xiao, J.}
\newblock 2014.
\newblock 3d shapenets for 2.5d object recognition and next-best-view
  prediction.
\newblock {\em CoRR abs/1406.5670\/}.

\bibitem[\protect\citename{Xu et~al\mbox{.} }2011]{Liefei2011}
{\sc Xu, L., Quynh~Dinh, H., Mordohai, P., and Ramsay, T.}
\newblock 2011.
\newblock Detecting patterns in vector fields.
\newblock {\em 49th AIAA Aerospace Sciences Meeting including the New Horizons
  Forum and Aerospace Exposition}, 6.

\bibitem[\protect\citename{Zhao et~al\mbox{.}
  }2014]{Zhao:2014:ISU:2631978.2574860}
{\sc Zhao, X., Wang, H., and Komura, T.}
\newblock 2014.
\newblock Indexing 3d scenes using the interaction bisector surface.
\newblock {\em ACM Trans. Graph. 33}, 3, 22:1--22:14.

\bibitem[\protect\citename{Zheng et~al\mbox{.} }2013]{CGF:CGF12039}
{\sc Zheng, Y., Cohen-Or, D., and Mitra, N.~J.}
\newblock 2013.
\newblock Smart variations: Functional substructures for part compatibility.
\newblock {\em Computer Graphics Forum 32}, 2pt2, 195--204.

\bibitem[\protect\citename{Zhu et~al\mbox{.} }2014]{Zhu2014}
{\sc Zhu, Y., Fathi, A., and Fei-Fei, L.}
\newblock 2014.
\newblock Reasoning about object affordances in a knowledge base
  representation.
\newblock In {\em Computer Vision – ECCV 2014}, D.~Fleet, T.~Pajdla,
  B.~Schiele, and T.~Tuytelaars, Eds., vol.~8690 of {\em Lecture Notes in
  Computer Science}. Springer International Publishing, 408--424.

\end{thebibliography}
\newpage
\section*{Appendix}

\label{app:attributes}
\changed{
We analyze the captured motion flows by computing a set of first-order attributes on a vector field. As shown in Xu et al.~\shortcite{Liefei2011} the gradient of a 3D vector field $v_i$ is an asymmetric 3D tensor field that can be decomposed into symmetric $S$ and asymmetric $A$ components. While the diagonal and off-diagonal entries of $S$ represents dilatation and shearing respectively, the off-diagonal entries of $A$ measure vorticity. Given the tensor $T(p)$ of a point $p\in v_i$, the decomposition is defined as   
{\small
\[
\boldsymbol{T(p)} = 
\begin{pmatrix}
    T_{11} & T_{12} & T_{13}  \\
    T_{21} & T_{22} & T_{23}  \\
    T_{31} & T_{32} & T_{33}  \\
\end{pmatrix}
= 
\begin{pmatrix}
    \frac{\partial V_x}{\partial x} & \frac{\partial V_x}{\partial y} & \frac{\partial V_x}{\partial z}  \\[2mm]
    \frac{\partial V_y}{\partial x} & \frac{\partial V_y}{\partial y} & \frac{\partial V_y}{\partial z}  \\[2mm]
    \frac{\partial V_z}{\partial x} & \frac{\partial V_z}{\partial y} & \frac{\partial V_z}{\partial z}  \\[2mm]
\end{pmatrix}
= \boldsymbol{A + S},
\]} 
with $S$ and $A$ given by
{\small
\[
\boldsymbol{S} = 
\begin{pmatrix}
    \epsilon_1 & \frac{1}{2}\theta_3 &  \frac{1}{2}\theta_2\\[2mm]
    \frac{1}{2}\theta_3 & \epsilon_2 &  \frac{1}{2}\theta_1\\[2mm]
    \frac{1}{2}\theta_2 &  \frac{1}{2}\theta_1 & \epsilon_3\\
\end{pmatrix}
\boldsymbol{~~A} = 
\begin{pmatrix}
    0          & -\omega_3 & \omega_2  \\[2mm]
    \omega_3   &         0 & -\omega_1  \\[2mm]
    -\omega_2  & \omega_1  & 0  \\
\end{pmatrix}, 
\]}where $\epsilon_i$ and $\theta_i$ represent the diagonal and off-diagonal components of $S$:  
{\small
\[
	\epsilon_1 = \frac{\partial V_x}{\partial x}, \epsilon_2 = \frac{\partial V_y}{\partial y}, \epsilon_3 = \frac{\partial V_z}{\partial z}
\]
\[
	\theta_1 = \frac{\partial V_z}{\partial y} + \frac{\partial V_y}{\partial z}, 
	\theta_2 = \frac{\partial V_x}{\partial z} + \frac{\partial V_z}{\partial x}, 
	\theta_3 = \frac{\partial V_y}{\partial x} + \frac{\partial V_x}{\partial y}, 
\]
}
and $\omega_i$ the off-diagonal entries of $A$: 
{\small
\[
	\omega_1 = \frac{1}{2}(\frac{\partial V_z}{\partial y} - \frac{\partial V_y}{\partial z}), 
	\omega_2 = \frac{1}{2}(\frac{\partial V_x}{\partial z} - \frac{\partial V_z}{\partial x}),
	\omega_3 = \frac{1}{2}(\frac{\partial V_y}{\partial x} - \frac{\partial V_x}{\partial y}).
\]}Based on this formulation, the attributes tensor magnitude $\boldsymbol{M_t}$, dilatation magnitude $\boldsymbol{M_d}$, magnitude of shear strain rate $\boldsymbol{M_s}$ and vorticity magnitude $\boldsymbol{M_{\overrightarrow{\omega}}}$ are given by 
\begin{table}[h]
\begin{center}
	\begin{tabular}{ll}
		\small$\boldsymbol{M_t} = \sqrt{\frac{1}{2} \sum T_{i,j}^2},$ & \small$\boldsymbol{M_d~~} = \sqrt{\sum \epsilon_{i}^2},$\\[5mm]

		\small$\boldsymbol{M_s} = \sqrt{\sum \theta_{i}^2},$ & \small$\boldsymbol{M_{\overrightarrow{\omega}}} = \sqrt{\sum \theta_{i}^2}$, \\
	\end{tabular}	
\end{center}
\end{table}
with $i, j=1, 2, 3$. Additionally, we derive the attributes vector orientation $\boldsymbol{O}$, and vector magnitude $\boldsymbol{M}$ directly from the vector field. In contrast to the approach of Xu et al.~\shortcite{Liefei2011}, we do not consider the attribute pressure $\boldsymbol{P}$, as it cannot easily be acquired for all motion drivers in our framework. }

\end{document}